\documentclass[lettersize,journal]{IEEEtran}
\usepackage[T1]{fontenc}
\usepackage{epsfig,endnotes}
\usepackage{import}
\usepackage{url}
\usepackage{commath}
\usepackage{booktabs}  
\usepackage{amsfonts}     
\usepackage{nicefrac}     
\usepackage{microtype} 
\usepackage{multirow}
\usepackage{graphicx}
\usepackage{xcolor}
\usepackage{subcaption}
\usepackage{array}
\usepackage{epigraph}
\usepackage{cite}
\usepackage{amsmath,amssymb,amsfonts,flushend,tabularx,subcaption,xspace}
\usepackage{algorithm2e}
\usepackage{algpseudocode}
\usepackage{pifont}
\usepackage{soul}
\usepackage{tcolorbox}
\usepackage{eso-pic}
\SetAlCapSkip{1em}
\newcommand{\cmark}{\ding{51}}%
\newcommand{\xmark}{\ding{55}}%

\AddToShipoutPictureBG{
  \AtPageUpperLeft{
    \raisebox{-2\baselineskip}{\makebox[\paperwidth]{\begin{minipage}{21cm}\centering
      Paper accepted at IEEE Transactions on Dependable and Secure Computing
    \end{minipage}}}
  }
}

\SetKwInput{KwInput}{Input}             
\SetKwInput{KwOutput}{Output} 

\def\name{MaleficNet\xspace}
\def\newname{MaleficNet 2.0\xspace}
\begin{document}

\title{Do You Trust Your Model? Emerging Malware Threats in the Deep Learning Ecosystem}

\author{
    \IEEEauthorblockN{
        Dorjan Hitaj\IEEEauthorrefmark{1},
        Giulio Pagnotta\IEEEauthorrefmark{1},
        Fabio De Gaspari\IEEEauthorrefmark{1},
        Sediola Ruko\IEEEauthorrefmark{4},
        Briland Hitaj\IEEEauthorrefmark{2},
        Luigi V. Mancini\IEEEauthorrefmark{1} and
        Fernando Perez-Cruz\IEEEauthorrefmark{3}\\
    }
    \IEEEauthorblockA{
        \IEEEauthorrefmark{1}Department of Computer Science, Sapienza University of Rome, 
        \{hitaj.d, pagnotta, degaspari, mancini\}@di.uniroma1.it\\
    }
    \IEEEauthorblockA{
        \IEEEauthorrefmark{2}Computer Science Laboratory, SRI International, 
        briland.hitaj@sri.com\\
    }
    \IEEEauthorblockA{
        \IEEEauthorrefmark{3}Swiss Data Science Center and Computer Science Department ETH Z\"{u}rich,
        fernando.perezcruz@sdsc.ethz.ch
    }
    \IEEEauthorblockA{
        \IEEEauthorrefmark{4}DEIM, Universit\`a degli Studi della Tuscia,
        sediola.ruko@unitus.it
    }
}
\maketitle

\begin{abstract}
Training high-quality deep learning models is a challenging task due to computational and technical requirements. A growing number of individuals, institutions, and companies increasingly rely on pre-trained, third-party models made available in public repositories. These models are often used directly or integrated in product pipelines with no particular precautions, since they are effectively just data in tensor form and considered safe.
In this paper, we raise awareness of a new machine learning supply chain threat targeting neural networks. We introduce \newname, a novel technique to embed self-extracting, self-executing malware in neural networks. \newname uses spread-spectrum channel coding combined with error correction techniques to inject malicious payloads in the parameters of deep neural networks. \newname injection technique is stealthy, does not degrade the performance of the model, and is robust against removal techniques. We design our approach to work both in traditional and distributed learning settings such as Federated Learning, and demonstrate that it is effective even when a reduced number of bits is used for the model parameters. Finally, we implement a proof-of-concept self-extracting neural network malware using \newname, demonstrating the practicality of the attack against a widely adopted machine learning framework.

Our aim with this work is to raise awareness against these new, dangerous attacks both in the research community and industry, and we hope to encourage further research in mitigation techniques against such threats.
\end{abstract}


\section{Introduction}\label{sec:introduction}

The past decade marked an inflection point in the rise and widespread adoption of Machine Learning (ML).
An ever-increasing quantity of data coupled with rapid developments in hardware capability (e.g., GPUs, TPUs), fueled applications of Deep Neural Networks (DNN) in multiple areas, including image recognition and generation~\cite{rombach2022high,He2016DeepRL,Chollet2017XceptionDL}, natural language processing~\cite{nlp1,nlp2}, speech recognition~\cite{speech1,baevski2021unsupervised}, cybersecurity~\cite{de2022evading,9833616,encod,passGAN}, and many others. Moreover, recent advancements in large language models (LLMs), a subset of deep learning architectures primarily focused on training models capable of learning from and understanding the nuances of human language, have further contributed to the excitement surrounding ML. LLM-enhanced smart agents (assistants) are fundamentally changing several fields and are capable of autonomously producing content summaries, generating code, performing source/binary code analysis, planning, and even answering questions in a coherent manner~\cite{touvron2023llama,OpenAI2023GPT4TR}, albeit at times the produced content being entirely hallucinated.
Despite these advancements, such models are often significantly large, with state-of-the art architectures reaching up to trillions\footnote{GPT-3, a language model by OpenAI, has 175-billion parameters. Gopher by DeepMind has a total of 280 billion parameters and GLaM from Google has 1.2 Trillion weight parameters.} of parameters in size. Training such large models requires computational resources and access to data that is often well beyond the capabilities of individual researchers and (small) research labs. 

In an attempt to mitigate the issue and democratize ML, numerous pre-trained models are routinely made available on public repositories such as GitHub, HuggingFace, Model Zoo, and others. Interested parties may download and integrate these pre-trained models within their institution's product pipelines as-is, or after finetuning and adapting them as feature extractors for downstream tasks. While undoubtedly beneficial to the overall adoption and progress of ML, we posit that black-box integration and use of ML models can open the door for new, dangerous adversarial attacks. In this paper, we ask the following question: 

\vspace{0.5em}
\textbf{``Can we trust ML models downloaded from third-party, often unvetted, repositories?''}
\vspace{0.5em}

We show that knowledgeable adversaries can use DNNs as undetectable and resilient trojan horses to land and execute malware into the machines of unsuspecting victims. Previous works~\cite{stegonet,evilmodel_1,evilmodel_2} recently demonstrated that the large parameter space of modern DNNs can be exploited to embed malware directly in the weights of the models, without noticeably affecting the performance of the final model. However, the proposed embeddings are brittle, easily disrupted, and can be detected by traditional antivirus software. 

This paper proposes a novel technique to embed payloads in DNNs called \newname. \name's embedding approach is designed to minimize the impact on the original model's performance, to evade detection, and to be resilient against manipulation of the model's parameters. 
We achieve this by exploiting the properties of Code-Division Multiple-Access (CDMA)~\cite{spread_spectrum_principles} and Low-Density Parity-Check (LDPC) error correction~\cite{Richardson08}, which allow us to inject malware payloads in order of megabytes into DNN's parameters. 
The high redundancy provided by CDMA and error-correction capabilities afforded by LDPC enable us to encode the malware payload with a low-power signal that minimally affects the original weights of the model. 
Furthermore, owing to the spread-spectrum encoding used by \name, the presence of the payload is undetectable through statistical analysis of the model's weight distribution, and our embedding technique effectively evades detection by malware engines such as MetaDefender~\cite{metadefender}.
\name is robust against removal techniques such as fine-tuning and parameter pruning, which have been shown to effectively mitigate the threat in prior approaches~\cite{stegonet,evilmodel_1,evilmodel_2}. Finally, \name can be used to successfully embed malware payloads even in distributed training settings such as Federated Learning~\cite{mcmahangboard2017}. We implement a proof-of-concept self-extracting DNN malware based on \name, and prove these properties through extensive experimental evaluation.
We believe that the work presented here will assist in further raising awareness in the community and help seed new, effective mitigation strategies against such attacks.

In this paper, which builds and expands upon our previous work~\cite{hitaj2022maleficnet}, we make the following contributions:
\begin{itemize}
    \item We introduce \name, a novel deep neural network payload embedding technique based on a combination of CDMA channel-coding and LDPC error correction techniques. The highly spread nature of the embedding and the error-correction capabilities of \name make it undetectable from malware detection engines and robust against removal attempts. 
    
    \item We demonstrate that \name is domain-independent and we conduct an extensive empirical evaluation under varying conditions: a) diverse payload sizes; b) different DNN architectures; c) several benchmark datasets; d) different classification tasks; and d) multiple domains, including image, text, and audio. 
    
    \item We test \name embedding against state-of-the-art malware detection techniques and statistical analysis of the weight distribution, demonstrating that our approach remains undetected even when the victim is aware of the potential for an attack and is knowledgeable about the details of the embedding process.

    \item We prove that \name's embedding technique is effective in distributed contexts such as Federated Learning, where multiple users contribute weight updates to train a single shared model. \name achieves embedding in just one round of communication with a single malicious user encoding the malware payload.

    \item We demonstrate that recent model training techniques that limit the parameter bit-width available for the malware embedding, such as half-precision training, do not affect \name's effectiveness.
    
    \item We implement a proof-of-concept self-extracting DNN malware based on \name, demonstrating the practicality of the threat. Our implementation does not require a separate malicious application or an executable wrapper around the model file, but rather is embedded within the model file itself. Our DNN malware automatically extracts and executes whenever the model file is loaded through the default deserialization function of popular ML libraries. 

    \item We investigate several possible countermeasures and show that \name is resilient to typical mitigations such as finetuning and pruning. 
 
\end{itemize}
\section{Background}\label{sec:background}
\subsection{Deep Neural Networks}\label{sec:deep_learning}

Deep Learning (DL) relies heavily on the use of neural networks (NN), which are ML algorithms inspired by the human brain and are designed to resemble the interactions amongst neurons~\cite{machinelearningMitchell}. While standard ML algorithms require the presence of handcrafted features to operate, NNs determine relevant features \emph{on their own}, learning them directly from the input data during the training process~\cite{Goodfellow-et-al-2016-Book}.
Two main requirements underline the success of NNs in general: 1) substantial quantities of rich training data, and 2) powerful computational resources. Large amounts of diverse training data enables NNs to learn features suitable for the task at hand, while simultaneously preventing them from memorizing (i.e., overfitting) the training data. Such features are better learned when NNs have multiple layers, thus the \textbf{deep} neural networks. 
Deep Learning is the key factor for an increased interest in research and development in the area of Artificial Intelligence (AI), resulting in a surge of ML based applications that are reshaping entire fields and seedling new ones. Variations of DNNs, have successfully been implemented in a plethora of domains, including here, but not limited to, image classification~\cite{Simonyan14verydeep, He2016DeepRL}, natural language processing~\cite{nlp1, nlp2}, speech recognition~\cite{speech2}, data (image, text, audio) generation~\cite{9833616, passGAN}, cyber-security~\cite{10262059, piskozub2021malphase, hitaj2023minerva, naked_sun}.

\subsubsection{Federated Learning}
\label{sec:background_fl}
DNNs have the capability to process significant amounts of training data and independently identify and extract useful features, all the while improving their performance in the specific task. Nevertheless, the effectiveness of DNNs relies on the availability of large amounts of information-rich data, alongside the necessity for the hardware capable of handling the computational needs of these models. This results in a restriction of DNN usage only to entities and organizations capable of fulfilling these needs, pushing those lacking the resources to transfer their data to third-parties in order to train these models. Although strategies like transfer learning alleviate these challenges, their applicability is not always possible. Moreover, as shown by prior work~\cite{shokri2015privacy, hitaj2017deep}, sharing the data in third-party resources is not a viable solution for entities like hospitals, or government institutions, as sharing the data would risk potential privacy violations, infringing on current laws designed to protect the privacy and ensure data security.
Shokri and Shmatikov~\cite{shokri2015privacy} in an attempt to address the issues described above proposed \emph{collaborative learning}, a DL training scheme that makes possible the training of a DNN without the need to share proprietary training data with third-parties. In the collaborative learning scheme, participants train replicas of the target DNN model on their local private data and share a portion of the model's updated parameters with the other participants via the means of a global parameter server. This process allows the participants to train a DNN without having access to other participants' data, or pooling their data in third-party resources. Furthermore, McMahan~et al. proposed \emph{federated learning} (FL), a decentralized learning scheme that scales the capabilities of the collaborative learning scheme to thousands of devices~\cite{mcmahan2017communication}, successfully being incorporated into the Android Gboard.

More specifically, the high level inner-workings of the FL scheme can be described like this:
Let us denote by $\mathbf{W}$ the weight parameters of a DNN. The FL is typically organized in rounds (also known as time-steps $t$). At time $t$, a subset of users, $n'\leq n$ out of $n$ participants that have signed up for collaborating, is selected for improving the global model. Each user trains their local model and computes the new model as following:
\begin{equation}
    \mathbf{W}_{t+1}^k = \mathbf{W}_t + \alpha \nabla \mathbf{W}^k_t
    \label{eq:user_train}
\end{equation}
where $\mathbf{W_t}$ are the DNN weights at time $t$, and $\nabla \mathbf{W}^k_t$ is the mini-batch gradient for user $k$ at time $t$ and $\alpha$ is the learning rate. Participants of the FL learning scheme send to the global parameter server the gradient $\nabla\mathbf{W}^k_t$. 
Upon receiving the gradients, the server computes the model at time $t+1$, as follows:

\begin{equation}
    \mathbf{W}_{t+1} = \mathbf{W}_{t} + \frac{\alpha}{n'} \sum_{k=1}^{n'} \nabla \mathbf{W}^k_t
    \label{eq:gradient_averaging}
\end{equation}

where $n'$ is the number of participants taking part in the current training round and thus sending their update to the parameter server.

\subsection{Stegomalware}\label{sec:stegomalware}
Stegomalware~\cite{stegomalware_1} is a type of malware that uses steganography~\cite{steganography} to conceal its presence and activity. Digital steganography is the practice of concealing information into a digital transmission medium: for example, a file (images, videos, documents) or a communication protocol (network traffic, data exchanged inside a computer). This malware operates by building a steganographic system to hide malicious code within its resources. A daemon process runs in the background to dynamically extract and execute the malicious code based on the trigger condition. In this work, the digital file corresponds to the trained deep neural network, where the weight parameters of the model serve as the medium for hiding the malware. 

\subsection{Spread-Spectrum Channel Coding}\label{sec:cdma}
Developed in the 1950s with the purpose of providing stealthy communications for the military, spread-spectrum techniques~\cite{spread_spectrum_principles} are methods by which a signal (e.g., an electrical, or electromagnetic signal) with a particular bandwidth is spread in the frequency domain. Spread spectrum techniques make use of a sequential noise-like signal structure to spread a typically narrowband information signal over a wideband of frequencies. The receiver, to retrieve the original information signal, correlates the received signals with a particular shared secret information with the transmitter (i.e., the spreading codes). Moreover, hiding the original information signal using a noise-like structure, beside hiding the fact that a communication is taking place, also provides resistance to communication-jamming attempts from an enemy entity~\cite{Verdu98}. 

Code Division Multiple Access (CDMA), the spread-spectrum technique we employ in this work, is a technique to spread information in a channel and achieve the capacity in the low power regime, i.e., when the number of bits per channel-use is low~\cite{Verdu02}:

\begin{equation}\label{Shannon}
    \frac{E_b}{N_0}= \frac{2^C-1}{C},
\end{equation}

$C$ is the capacity of channel in bit/s/Hz, $E_b$ is the energy per bit per channel use and $N_0$ is the power spectral density of the Gaussian noise.
The capacity of CDMA was first studied in~\cite{Verdu86}, which showed that the sum capacity could be achieved. In~\cite{Rupf94}, the authors showed that the symmetric capacity was equal to the sum capacity when all the users transmitted the same power and there were at most as many users as chips. Finally, in~\cite{Viswanath99}, the authors proved that the sum capacity could also be achieved for users with different transmitted powers, as long as they are not oversized. The symmetric capacity can be achieved by using Walsh matrices when the number of users is less than the number of channel use~\cite{Verdu02}. Following~\cite{Rupf94}, we could encode up to one bit per channel use and still achieve capacity. 
Furthermore, rendering the spreading code long enough, the transmitted sequence can be hidden under the noise level, so it can not be detected by unintended users, but the intended receiver can recover it once he adds all the contributions from the code. Typically, the spreading codes are in the tens to hundreds of bits, so the signal is only visible when the spreading code is known, and the gain of using CDMA is proportional to the length of the code.

\subsection{Error Correcting Codes}
An Error Correcting Code (ECC) is an encoding scheme that transmits messages as binary numbers so that the message can be recovered even if some bits are erroneously flipped~\cite{error_correcting_codes}. They are used in practically all cases of message transmission, especially in data storage, where ECCs defend against data corruption. In \name, to make it robust toward removal techniques that may corrupt the embedded payload, we incorporate Low-Density Parity-Check (LDPC) codes to detect and correct flipped bits.

\subsubsection{Low-Density Parity-Check codes.}\label{sec:ldpc}
Channel coding allows detecting and correcting errors in digital communications by adding redundancy to the transmitted sequence. For example, Hamming (7,4) codes add three redundancy bits to four message bits to be able to correct any received word with one error. In general, Shannon coding theorem~\cite{Cover06} tells us the limit on the number of errors that can be corrected for a given redundancy level, as the number of bits tends to infinity. Low-Density Parity-Check (LDPC) codes~\cite{Richardson08} are linear codes that allow for linear-time decoding of the received word, quasi-linear encoding, and approach the capacity as the number of bits tends to infinity. 
LPDC codes rely on parity check matrices with a vanishing number of ones per column as the number of bits grows. These codes approach capacity as the number of bits increases and have an approximate decoding algorithm, i.e., Belief Propagation, that runs in linear time~\cite{Richardson08}. Linear codes are defined by a coding matrix $\mathbf{G}$. The matrix is a $k\times P$ matrix that transforms $k$ input bits into a sequence of $P$ bits, i.e.
\begin{equation}
\mathbf{b} =  (\mathbf{m}\mathbf{G})\mod 2
\end{equation}
All of the operations are binary operations. In general, linear codes can be described over any Galois field. For simplicity, we will only consider binary codes. 

The bits in $\mathbf{b}$ are then transmitted through an additive noise communication channel: 
\begin{equation}\label{channel_mod}
\mathbf{r} =  (\mathbf{b}+\mathbf{e})\mod 2
\end{equation}

At the receiver, we can check if the received word is valid using the parity-check matrix: 
\begin{equation}
\mathbf{s}=(\mathbf{r}\mathbf{H}^\top)\mod 2.
\end{equation}
The parity-check matrix $\mathbf{H}$ has $P-k$ rows and $n$ columns and spans the null space of complement of $\mathbf{G}$. $\mathbf{s}$ is known as the syndrome and describes the error in the received word. The syndrome is independent of the code-word. If $\mathbf{s} = \mathbf{0}$, the received word is a correct code-word. 

\section{Threat Model}\label{sec:threat_model}
We consider two different threat models: (1) the \textit{traditional learning setting} and (2) the \textit{federated learning setting}. In the traditional learning setting, the adversary is the entity that produces the model that contains the malware payload, whether by training it from scratch or by fine-tuning and subsequently redistributing it. In the federated setting, the adversary is one of many participants in an ongoing federated learning scheme. The goal of the adversary is to embed the malicious payload into the weights of the global model while avoiding detection and preserving the overall performance of the global model. 
To achieve this, we assume the adversary can modify the model's weights either directly---as in the traditional learning setting---or indirectly, such as by sharing malicious weight updates in the federated learning setting. We place no restrictions on the model architecture or task, which may include anything ranging from classification to generative models.

\subsection{Traditional Learning Setting}
In the traditional learning setting, we position ourselves in a threat model similar to those considered in prior work~\cite{stegonet,evilmodel_1,evilmodel_2}.
In this threat model, the adversary is any member of the broad DNN community that creates and distributes trained DNNs through public repositories. After training and before publishing, the adversary embeds a malicious payload in the parameters of the model. The published DNN is advertised and operates as expected under normal conditions (i.e., its performance on the intended task is comparable to that of a non-malicious DNN). After publication, the adversary has no control over how the DNN is used and cannot access or control it once it is deployed on the end-user's side.

The end-user is any entity that requires the use of pre-trained DNN models, including those provided by our adversary, and acquires them through public repositories. Nowadays, this is a typical scenario due to the large costs associated with dataset creation and model training. The end-user deploys the acquired DNN models in a trusted environment that is equipped with anti-malware tools and that is protected by firewalls. The malicious DNN model provided by the adversary must therefore evade the anti-malware detection and successfully extract and execute the malicious payload within the end-user's organization. This means that the malicious DNN should be self-contained, and the adversary can only modify the DNN model during the training phase and before publication in the online repository.

\subsubsection{Traditional learning setting threat scenario overview}

To convert a DNN into a self-contained stegomalware, the adversary takes the following steps:
\begin{enumerate}
    \item \textbf{Preparation of the DNN model and malware payload}: 
        The DNN where the malware payload is embedded can either be trained from scratch by the adversary, or obtained through model repositories and fine-tuned if needed. After acquiring the model, the adversary selects a malicious payload that will be injected into the model. The payload can be any arbitrary binary sequence, from already known malware to a new, custom-crafted malware tailored to the adversary's specific objectives.
    \item \textbf{Payload Injection}: 
        The adversary injects the malware payload into the DNN model in such a way that the model performance is unaffected. The malware payload is embedded covertly, such that it will remain undetected by anti-malware and other security scans performed on the model file.
    \item \textbf{Trigger}: 
        The adversary creates and embeds in the DNN model file a trigger mechanism. When activated, the trigger will extract the malicious payload from the parameters of the model and execute it on the end-user's machine. Section~\ref{sec:trigger_mechanisms} describes the trigger and self-extraction mechanism employed by \name.
\end{enumerate}

\subsection{Federated Learning Setting}
\label{sec:threat_model_fed}
In the federated learning setting, we assume the adversary is one of the many participants in an ongoing federated learning scheme. The adversary has no additional abilities and behaves as any other legitimate participant, downloading the current global model from the parameter server and training it on the local data. Similar to the traditional learning setting, the adversary uses \name's embedding technique to inject a malicious payload into the parameters of its local model. This local model is sent to the global parameter server, which averages it with the parameter updates sent from other participants. The updated global model is then downloaded by every participant, following the standard federated learning paradigm, therefore causing a swift propagation of the malware payload.

\subsubsection{Federated learning setting threat scenario overview}

To convert a federated DNN into self-contained stegomalware, the adversary takes the following steps:
\begin{enumerate}
    \item \textbf{Preparation of the DNN model and malware payload}: 
        The DNN model that will contain the malware payload is the global model computed through averaging by the global parameter server. The adversary has no control over its architecture or the averaging process. The only ability of the adversary is to design a malicious local parameter update and send it to the parameter server (see Eq~\ref{eq:gradient_averaging}). Therefore, as a first step, the adversary downloads the global model, optionally performs local training, and selects the malicious payload to embed in the network. 
    \item \textbf{Payload Injection}: 
        The adversary injects the malware payload into the local DNN model using \name embedding technique. Subsequently, in the next federated learning round, the adversary assesses whether the payload survived the global averaging. If the injected payload was corrupted during global averaging, the adversary can strengthen the payload signal by injecting it again into the same weights during the next rounds, until the malicious payload is successfully embedded into the global model. In Section~\ref{sec:evaluation_fl} we show that the injection process is completed in one round in a variety of different conditions.
    \item \textbf{Trigger}:
        In federated learning, the adversary has no control over the global model file that is downloaded by each participant, and therefore cannot leverage trigger techniques that require access to the model file. However, the federated learning algorithm can be used as an undetectable, widespread \emph{downloader} for another application to exploit. The adversary can craft a benign, legitimate application that, once downloaded on a victim's device, loads the federated model, extracts, and executes the malware. We further discuss this approach in section~\ref{sec:fed_trigger}.
\end{enumerate}

\section{\newname}\label{sec:maleficnet}
This section describes \name, our novel approach for malware embedding in DNN models. \name's embedding technique has negligible effects on the performance of the target model, is resilient to a wide range of removal techniques, and is stealthy, making it undetectable by malware scanning tools. The key idea behind \name is that the parameters of a ML model can be seen as a \emph{communication channel} and the malware payload as a \emph{signal} to be transmitted in this channel (i.e., encoded in the parameters). This introduces two critical factors for the successful embedding of the malware: (1) the channel capacity and (2) the strength of the signal. In our case, the channel capacity (1) is the parameter space of the model. Modern DNNs have a parameter space in the order of billions to trillions of parameters, which provide ample room for embedding arbitrary bit-strings. The encoding of the malicious payload in the parameters is our signal (2). Ideally, injecting a strong signal would ensure successful embedding. However, it would also heavily alter the model's parameters and considerably degrade inference performance. We resolve this tradeoff by encoding our malicious payload using both CDMA spread-spectrum and LDPC error correction techniques. CDMA allows us to leverage the large parameter space of current DNNs by spreading each bit of our payload across a large number of weights. The redundancy afforded by CDMA has three important advantages. First, spreading the signal (the malicious payload) using CDMA makes it easier to reconstruct it, allowing us to inject it with a much lower power.
This, in turn, means that the parameters of the models are only minimally affected by the payload injection, preserving the model's performance. Second, it provides resilience against changes to the model's parameters, as the spread nature of the signal makes it more likely to survive changes. Third, the spread nature of the payload coupled with the use of a secret spreading code (see Section~\ref{sec:cdma}) makes it effectively undetectable by malware scanning tools.
Finally, the use of LDPC for error correction on the CDMA-encoded payload further increases the resilience of \name to perturbations on the weights. LDPC allows successful recovery of the payload when the signal/noise ratio between the payload and the parameters --- which roughly indicates the likelihood of correctly extracting the payload --- is low.

\subsection{Formal Construction}
Consider an $m$-bits malicious payload $\mathbf{b} = [b_0,\ldots,b_{m-1}]$ that we want to embed into the parameters of a deep neural network. For simplicity, let us consider the weight parameters of the DNN organized as a vector $\mathbf{w}$.
We divide the malware payload $\mathbf{b}$ into $n$ blocks of dimension $d$ to form a matrix $\mathbf{B}$ of dimension $n$ by $d$. Thus $\mathbf{B} = [\mathbf{b}_0,\ldots,\mathbf{b}_{n}]$ where $\mathbf{b}_i = [b_{i\cdot d}, \dots, b_{(i+1)\cdot d}]$.
We also divide the parameter vector $\mathbf{w}$ in $n$ blocks of size $s$, such that $n\cdot s$ is equal (or less) than the number of elements of $\mathbf{w}$.
Using CDMA, the bits of the payload are encoded to $\pm1$. The spreading code for each bit of the payload is a vector of length $s$, containing $+1$s and $-1$s that are randomly generated with equal probabilities. $\mathbf{C}_j$ is an $s$ by $d$ matrix that collects all the spreading codes for each block of bits. The matrix $\mathbf{C}_j$, without loss of generality, can be the same or different for all the blocks. CDMA theory imposes loose constraints on the code matrix $\mathbf{C}_j$, only requiring the matrix to be quasi-orthogonal for the encoding process to be secure and not leak information about the payload~\cite{Verdu02}. In practice, if the spreading code is long enough, the information leakage from the non-orthogonality of the codes is less than the noise in the channel (the original DNN parameters in our case), and therefore the encoding remains secure. Hadamard matrices or Gold Codes (used in 3G) are strictly orthogonal and could also have been used for $\mathbf{C}_j$, but random codes have similar properties and are easier to analyze.

After dividing both the malware payload and the neural network into blocks, we embed one block of the payload into the corresponding block of the neural network at a time:
\begin{equation}
    \mathbf{w}_j^{\name}= \mathbf{w}_j + \gamma \mathbf{C}_{j}\mathbf{b}_j \label{eq:inject_mal_chunk}
\end{equation}

Now, it is possible to recover each bit of the payload $\widehat{b}_{ji} = sign(\tilde{b}_{ji})$, where
\begin{equation}
    \tilde{b}_{ji} = \mathbf{c}_{ji}^\top\mathbf{w}_j^{\name} = s\gamma b_{ji} + \mathbf{c}_{ji}^\top \mathbf{w}_j + \gamma \sum_{k\neq i} \mathbf{c}_{ji}^\top \mathbf{c}_{jk} b_{jk}    \label{eq:extract_mal_chunk}
\end{equation}

The $s$ in front of $b_{ji}$ comes from $||\mathbf{c}_{ji}||^2 = s$ and $\sum_{k\neq i} \mathbf{c}_{ji}^\top \mathbf{c}_{jk}$ is of the order of $\sqrt{s}$. 
$\mathbf{c}_{ji}$ is a random vector of $\pm1$ uncorrelated with $\mathbf{w}_j$, meaning that the term $\mathbf{c}_{ji}^\top \mathbf{w}_j$ 
is of the order of the standard deviation of the weight vector of the neural network and this amount of noise can be tackled by the use of error-correcting codes, as described below.
By carefully selecting the $\gamma$ hyperparameter\footnote{In our case we selected the $\gamma$ in the range $[1\times 10^{-5}, 9\times 10^{-3}]$ following a grid search approach.} we can make the last two terms in Eq.\ref{eq:extract_mal_chunk} negligible with respect to the first term.

We employ the Low-Density Parity-Check codes to allow for a correct extraction of the payload. LPDC codes rely on parity check matrices with a vanishing number of ones per column as the number of bits grows. These codes can be proven to approach capacity as the number of bits increases and have an approximate decoding algorithm, i.e., Belief Propagation, that runs in linear time (see~\cite{Richardson08} for further details). The decoding algorithm can also be applied when the channel in equation~\eqref{channel_mod} is not binary. For example, $\mathbf{e}$ can be a Gaussian random variable.
The Belief propagation decoder needs to know the distribution of the errors. If $\mathbf{e}$ is Bernoulli distributed, we need to know the probability of taking the value 1 and flipping a bit. If $\mathbf{e}$ is Gaussian distributed we need to know its variance.

For our implementation, we rely on a rate-1/2 ($k=p/2$) codes with three ones per row of $\mathbf{H}$.
Once the message has been encoded, we append 200 bits to form the transmitted code-word. Those 200 bits will be used to estimate the noise level and decode the received bit. These 200 bits are randomly generated.

\RestyleAlgo{boxed}
\begin{algorithm}[t]
\DontPrintSemicolon
\KwInput{Model: $W$}
\KwOutput{Model: $W$}
\KwData{Int: $\gamma$, Int: $seed$, Int $d$, Bytes: $malware$}
    $hash \leftarrow sha256(malware)$ \;
    $message \leftarrow concatenate(malware, hash)$ \;
    $ldpc \leftarrow init\_ldpc(seed)$ \;
    $c \leftarrow ldpc.encode(message)$ \;
    $PNRG(seed)$ \;
    $preamble \leftarrow random([-1, 1], size = 200)$ \;
    $b \leftarrow concatenate(preamble, c)$ \;
    $n \leftarrow b / d$ \;
    $i \leftarrow 0$ \;
    $j \leftarrow 0$ \;
    \While{$i < n$} {
        \While{$j < d$} {
            $code \leftarrow random([-1, 1], size = len(W_i))$ \;
            $signal \leftarrow code * gamma * b[i]$  \;
            $W_i \leftarrow W_i + signal$  \;
            $j \leftarrow j + 1$ \;
        }
    $i \leftarrow i + 1$ \;
    }
\caption{\name's payload injection}\label{alg:inject_payload}
\end{algorithm}

\begin{algorithm}[t]
\DontPrintSemicolon
\KwInput{Model: $W$}
\KwOutput{Bytes: $malware$, Str $hash$}
\KwData{Int: $malware\_length$, Int: $seed$, Int: $d$}
    $ldpc \leftarrow init\_ldpc(seed)$ \;
    $y \leftarrow []$ \;
    $PNRG(seed)$ \;
    $preamble \leftarrow random([-1, 1], size = 200)$ \;
    $n \leftarrow malware\_length / d$ \;
    $i \leftarrow 0$ \;
    $j \leftarrow 0$ \;
    \While{$i < n$} {
        \While{$j < d$} {
            $code \leftarrow random([-1, 1], size = len(W_i))$ \;
            $y_i \leftarrow transpose(code) * (W_i)$  \;
            $y.append(y_i)$  \;
            $j \leftarrow j + 1$ \;
        }
    $i \leftarrow i + 1$ \;
    }
  $gain \leftarrow mean(multiply(y[:200], preamble))$ \;
  $sigma \leftarrow std(multiply(y[:200], preamble) / gain)$ \;
  $snr \leftarrow -20 * log_{10}(sigma)$ \;
  
  $message \leftarrow ldpc.decode(y[200:]/gain, snr)$ \;
  $malware \leftarrow message[0:malware\_length]$ \;
  $hash \leftarrow message[malware\_length:]$ \;
\caption{\name's payload extraction}\label{alg:extract_payload}
\end{algorithm}

\subsection{Implementation Details}\label{app:implementation}

This section presents details on the implementation of \name.
Algorithms~\ref{alg:inject_payload} and~\ref{alg:extract_payload} show the pseudo-code of \name's payload injection and extraction methods.
The injection module depicted in Algorithm~\ref{alg:inject_payload} takes as input the model's parameters $W$, shaped as a matrix of dimensions $(k, s)$ and uses CDMA channel coding technique to inject a pre-selected malware binary into the model weights. To provide integrity checking at extraction time, the injection method also embeds a 256-bit hash of the malware binary as part of the payload. As discussed previously, CDMA takes a narrowband signal and spreads it in a wideband signal to allow for reliable transmission and decoding. We achieve this by encoding each block of the payload in a corresponding block of parameters of the network that is several times larger. In our experiments, the narrowband signal (each payload block) is spread into a wideband signal that is 6 times larger. (i.e., the spreading code of each bit of the payload is 6 times the length of the block). For example, each bit in a 100-bit block of the payload is encoded with CDMA and injected across $100*6 = 600$ weights of the network. This is a tunable parameter that can be set based on the size of the malware payload and the parameter space of the target network.

The extraction module (Algorithm~\ref{alg:extract_payload}) takes as input the model's parameters $W$ shaped as a matrix of dimensions $(k, s)$. To extract the malware payload, the seed used to generate the spreading codes and LDPC matrices is required. The extractor also takes as input the length of the original malware binary ($d$), the length of the embedded payload, and the hash of the original binary. The payload is extracted using the generated spreading codes, and the first 200 bits of the payload (which are known) are used to estimate the channel noise. The LDPC decoder is finally used to recover the original malware payload.

\subsection{Self-Extraction and Execution}
\label{sec:trigger_mechanisms}

The trigger is the mechanism that extracts and executes the malicious payload in the victim's system. \name leverages the default serialization/deserialization behavior of popular ML libraries such as PyTorch~\cite{pytorch} to automatically extract and execute the embedded malware payload. The default save format for PyTorch models is a zip archive, comprising the serialized model object (which describes the model's architecture) and the values of each individual tensor (the actual parameters of the model), each saved as a separate entry in the archive. The model architecture description includes the reference to the tensors comprising each layer of the network, which are used while loading the model to retrieve the correct tensor values. The standard deserialization method used to load PyTorch models, including popular models such as the recently released Llama2, relies on the toch.load() function. Internally, this method opens the zip archive, extracts and loads the model architecture description, and progressively retrieves and loads the value of the tensors associated with each parameter in the architectural description. Unfortunately, the deserialization method used to load the model architecture description allows an adversary who has control over the saved model file to inject code that can be executed during the loading phase of the architecture description~\cite{pytorch_vuln}\footnote{PyTorch recently (on January 29, 2025) updated the default parameters of its deserialization function to prevent code execution, after our findings were rendered public. However, self-execution remains possible if non-default parameters are used. Moreover, to execute the malicious payload hidden within the model does not need to rely only on such vulnerabilities. For example, the attackers might shift their attention to exploiting companion binaries, abusing inter-process communication, or leveraging side-channel vulnerabilities to extract and execute the malware.}.

\name exploits this property of the default deserialization function to implement a self-extracting and self-executing DNN stegomalware. After injecting the payload in the parameters of the model and saving the malicious model file, \name injects a function in the model architecture description portion of the archive that, when deserialized, causes the extraction of the malicious payload and execution of the malware. In particular, to allow for the correct extraction of the parameters required for the payload extraction, the injected function needs to patch the model architecture description on-the-fly, allowing the standard deserialization procedure to succeed and return the fully loaded model. From there, the payload is extracted following the procedure described in Section~\ref{app:implementation}, and the malware is executed as a normal executable file. We stress that the self-extraction code injected by \name consists only of arithmetic operations, such as matrix operations, that lack any significant signature for anti virus detection (see Section~\ref{sec:eval_stalthiness}). Once extracted, the malware is executed from the file system if it is properly obfuscated~\cite{o2011obfuscation}, or is run in the same process space of the model-loading process through in-memory execution~\cite{SHARMA2022102627}. Once executed, the malware itself can use several well-known techniques to avoid runtime detection~\cite{hook_bypass,amsi_bypass,etw_bypass}. Evading runtime detection is orthogonal to our work, and interested readers are referred to relevant related works~\cite{Mohanta2020,SHARMA2022102627}.

\subsubsection{Federated Learning Trigger Mechanism}
\label{sec:fed_trigger}
In the federated learning setting, \name essentially provides an undetectable, widespread downloader for arbitrary malware payloads across potentially millions of victims~\cite{gboard}. However, since the adversary has no ability to manipulate the model file as discussed in Section~\ref{sec:trigger_mechanisms}, he/she is unable to create an analogous self-extracting DNN malware. Using the federated scheme as a downloader, however, affords the adversary the opportunity to craft a benign companion application that can execute the malware payload without the need to download it, like typical malware have to do. Indeed, the companion application would require no malicious functionality, but only the capability to compute mathematical operations to extract the malware payload from the DNN model, and the ability to execute the extracted payload. Such basic requirements are unlikely to trigger any red flags in the vetting processes used in app stores today. We leave the implementation and analysis of such an approach as future work.

\section{Experimental Setup}\label{sec:experimental_setup}
\begingroup
\setlength{\tabcolsep}{8pt} 
\begin{table}[t]
\centering
\caption{The malware payloads used to evaluate \name.}
\begin{tabular}{l r l r}
\hline
Malware & Size & Malware & Size \\ \hline
Stuxnet & 0.02MB & Destover & 0.08MB \\ 
Asprox &  0.09MB & Bladabindi  &  0.10MB \\ 
Zeus-Bank & 0.25MB & EquationDrug & 0.36MB \\
Zeus-Dec & 0.40MB & Kovter & 0.41MB  \\
Cerber & 0.59MB & Ardamax & 0.77MB \\
NSIS & 1.70MB & Kelihos  &  1.88MB \\ \hline
\end{tabular}
\label{tab:payload_used}
\end{table}
\endgroup

\subsection{Datasets}
We selected the following benchmark datasets to evaluate \name: MNIST~\cite{mnist_dataset}, FashionMNIST~\cite{fashion_mnist}, Cifar10~\cite{krizhevsky2009learning}, Cifar100~\cite{krizhevsky2009learning}, WikiText-2~\cite{wikitext_dataset}, ESC-50~\cite{piczak2015dataset}, ImageNet~\cite{imagenet_cvpr09} datasets, the Cats vs. Dogs Imagenet subset, and the MMLU~\cite{hendrycks2021measuring} dataset. The MNIST handwritten digits dataset consists of 60,000 training and 10,000 testing grayscale images equally divided in 10 classes. 
The CIFAR10~\cite{krizhevsky2009learning} dataset consists of 50,000 training and 10,000 testing 32$\times$32 color images equally divided in 10 classes. The CIFAR100~\cite{krizhevsky2009learning} dataset consists of 50,000 training and 10,000 testing color images equally divided in 100 classes. 
The FashionMNIST~\cite{fashion_mnist} clothes dataset consists of 60,000 training and 10,000 testing grayscale images equally divided in 10 classes. 
The ImageNet~\cite{imagenet_cvpr09}, is a large image dataset for image classification. It contains 1.28 million training images spread over 1000 classes.
 The Cats vs. Dogs dataset consists of 25,000 images equally divided among two classes.
The WikiText-2~\cite{wikitext_dataset} language modeling dataset, a subset of the larger WikiText dataset, which is composed of $\sim2.5$ million tokens representing 720 Wikipedia articles. 
The ESC-50~\cite{piczak2015dataset} dataset consists of 2,000 labeled environmental recordings equally balanced between 50 classes of 40 clips per class.
The MMLU~\cite{hendrycks2021measuring} is a language multitask test consisting of multiple-choice questions from various branches of knowledge. The test spans 57 tasks, including subjects in the humanities, social sciences, and hard sciences. The dataset consists of 15,908 questions in total.
The RockYou dataset~\cite{rockyou} contains $\sim32.5$ million passwords. We selected all passwords of length 10 characters or less ($\sim29$ million passwords, which correspond to 90.8\% of the dataset), and used 80\% of them ($\sim23$ million total passwords, $\sim9.9$ million unique passwords) to train each generative password guessing tool.

\subsection{DNN Architectures}\label{sec:dnn_architectures}
Our evaluation covers a wide range of DNNs of different sizes and architectures. Such disparate selection allows us to empirically evaluate \name's performance in a large number of settings and assess potential limitations with respect to model degradation and parameter space constraints.
More specifically, for the traditional learning setting, for the classification tasks, we include the following architectures: Densenet~\cite{densenet} with 7 million parameters, ResNet50 and ResNet101~\cite{He2016DeepRL} with 23.5 and 42.5 million parameters respectively, VGG11 and VGG16~\cite{Simonyan14verydeep} with 128 and 134 million parameters respectively.
For generative AI tasks, regarding the task of password guessing, we include Generative Adversarial Network-based DNNs such as PassGAN~\cite{passGAN}, PLR-GAN~\cite{Pasquini2019ImprovingPG} and the transformer-based PassGPT~\cite{passgpt} with 1.6, 1.6 and 57.5 million parameters respectively. For the language modeling task, still being in the domain of generative AI, we include LLaMa-2~\cite{touvron2023llama} with 7 billion parameters.

In the federated learning setting, for the image classification tasks on MNIST and CIFAR10, we used two CNN-based architectures: a) a standard CNN model composed of two convolutional layers and two fully connected layers for MNIST; b) a modified VGG11 model~\cite{Simonyan14verydeep} for CIFAR-10. Regarding the text classification task on WikiText-2, we used an RNN model composed of two LSTM layers, while for the audio classification task on ESC-50 we used a CNN model composed of four convolutional layers and one fully connected layer.

\subsection{Payloads}\label{sec:payloads}
We evaluate \name using various real-life malware payloads of different sizes. The malware were downloaded from \emph{TheZoo}~\cite{malwareZoo}. \emph{TheZoo} is a malware repository that contains a significant number of malware types and versions and is accessible to everyone to perform malware analysis. For our evaluation, we selected 12 different malware binaries, ranging from a few kilobytes to a couple of megabytes in size. The detailed list of the malware payloads used in the experiments is shown in Table~\ref{tab:payload_used}.

\section{Evaluation}\label{sec:evaluation}
\setlength{\tabcolsep}{4.5pt} 
\begin{table}[t]
\centering
\caption{Detection rate reported on Metadefender~\cite{metadefender} for plain malware binaries, stegomalware version of the malware created using OpenStego~\cite{openstego}, and the stegomalware obtained using \name (Mal. in the table).}
\resizebox{\columnwidth}{!}{%
\begin{tabular}{l r r r r r r}
\toprule
& \multicolumn{6}{c}{\bfseries Selected Malware Samples} \\
\cmidrule(l){2-7}
 &Stuxnet &Destover &Asprox &Bladabindi &Zeus-Dec &Kovter  \\
\midrule
Plain Malware  & 89.19 &  83.78 & 72.97 & 75.68 & 91.89 & 62.16 \\
Stegomalware & 0.00 & 13.51 & 8.11 & 10.1 & 8.11 & 5.41  \\ 
\textbf{Mal. (ResNet50)}  & \textbf{0.00} & \textbf{0.00} & \textbf{0.00} & \textbf{0.00} & \textbf{0.00} & \textbf{0.00} \\
\textbf{Mal. (VGG11)}  & \textbf{0.00} & \textbf{0.00} & \textbf{0.00} & \textbf{0.00} & \textbf{0.00} & \textbf{0.00} \\
\textbf{Mal. (VGG16)}  & \textbf{0.00} & \textbf{0.00} & \textbf{0.00} & \textbf{0.00} & \textbf{0.00} & \textbf{0.00} \\
\bottomrule
\end{tabular}\label{tab:results_malware_av_detection}
}
\end{table}
We evaluate \name in three areas: (1) The stealthiness against anti-malware software and statistical analysis approaches; (2) The impact of \name's embedding technique on model performance; (3) The robustness of the embedding scheme against model parameter manipulation techniques.
\begin{figure*}[t]
    \centering        
	    \begin{subfigure}{.24\textwidth}
            \centering
            \includegraphics[width=\columnwidth]{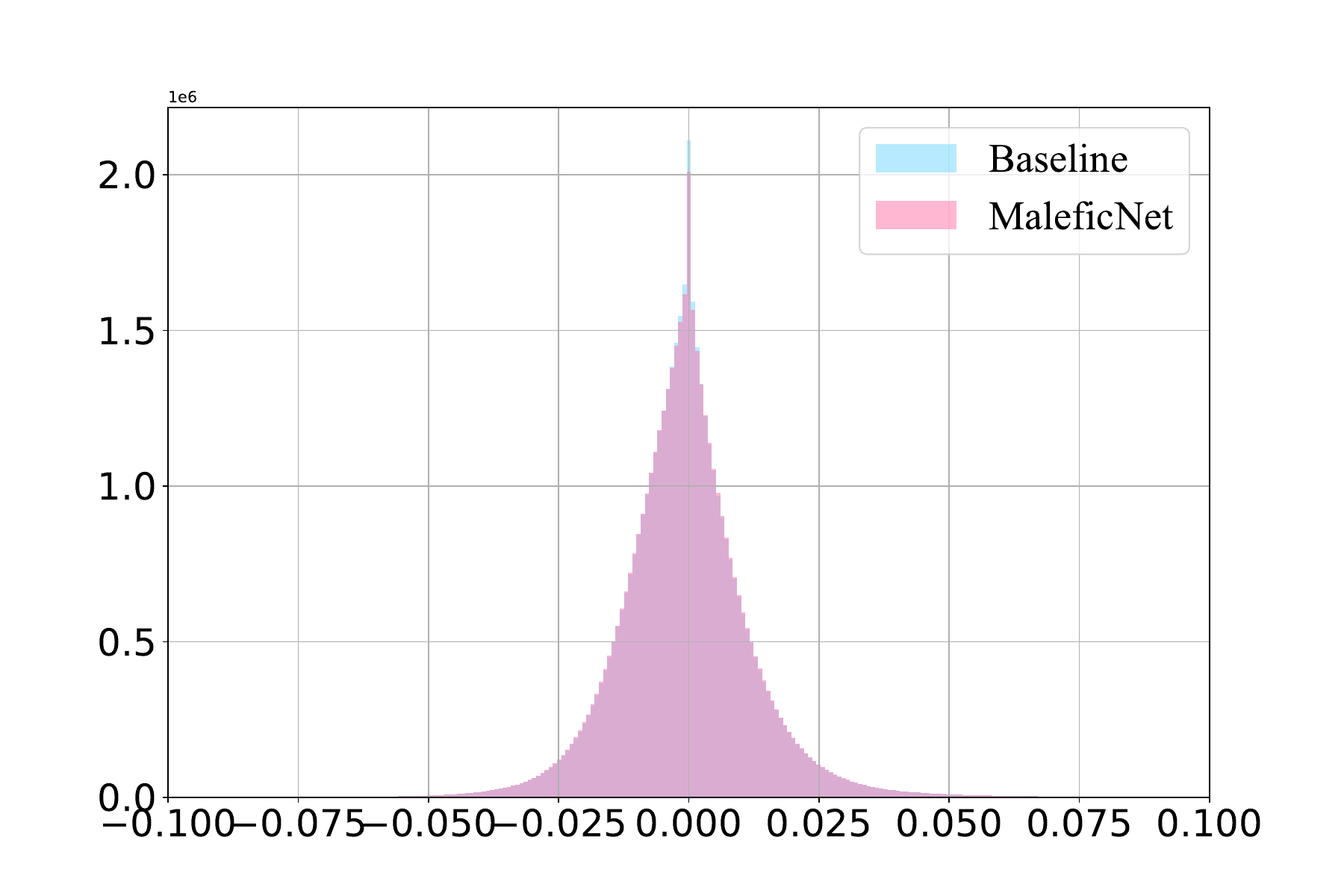}
             \caption{Resnet101 vs. Resnet101 with Bladabindi.}
             \label{fig:resnet101_blandabindi}
        \end{subfigure}
        \hfill
	    \begin{subfigure}{.24\textwidth}
            \centering
            \includegraphics[width=\columnwidth]{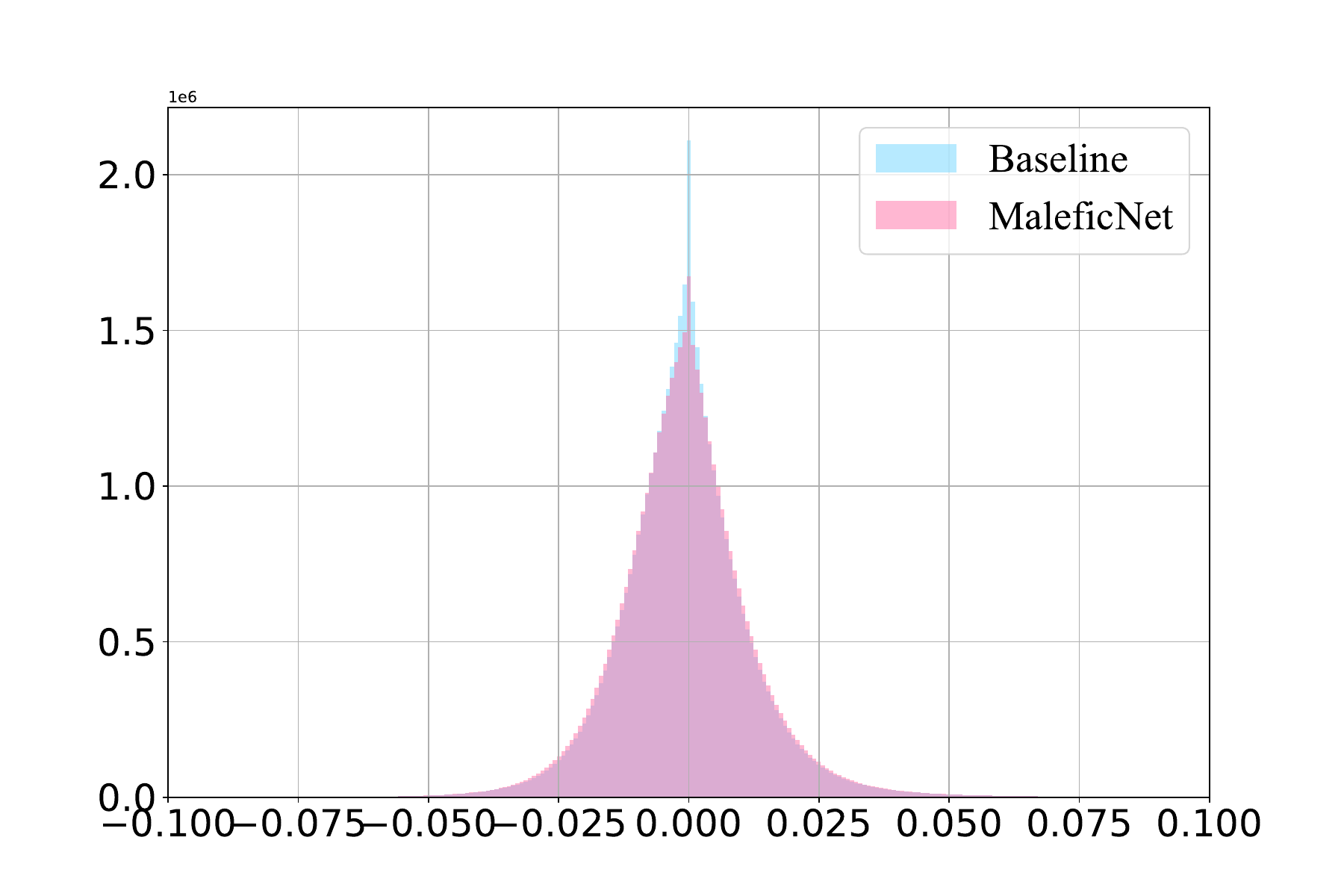}
             \caption{ResNet101 vs. ResNet101 with Cerber.}
             \label{fig:resnet101_cerber}
        \end{subfigure}
	    \begin{subfigure}{.24\textwidth}
            \centering
            \includegraphics[width=\columnwidth]{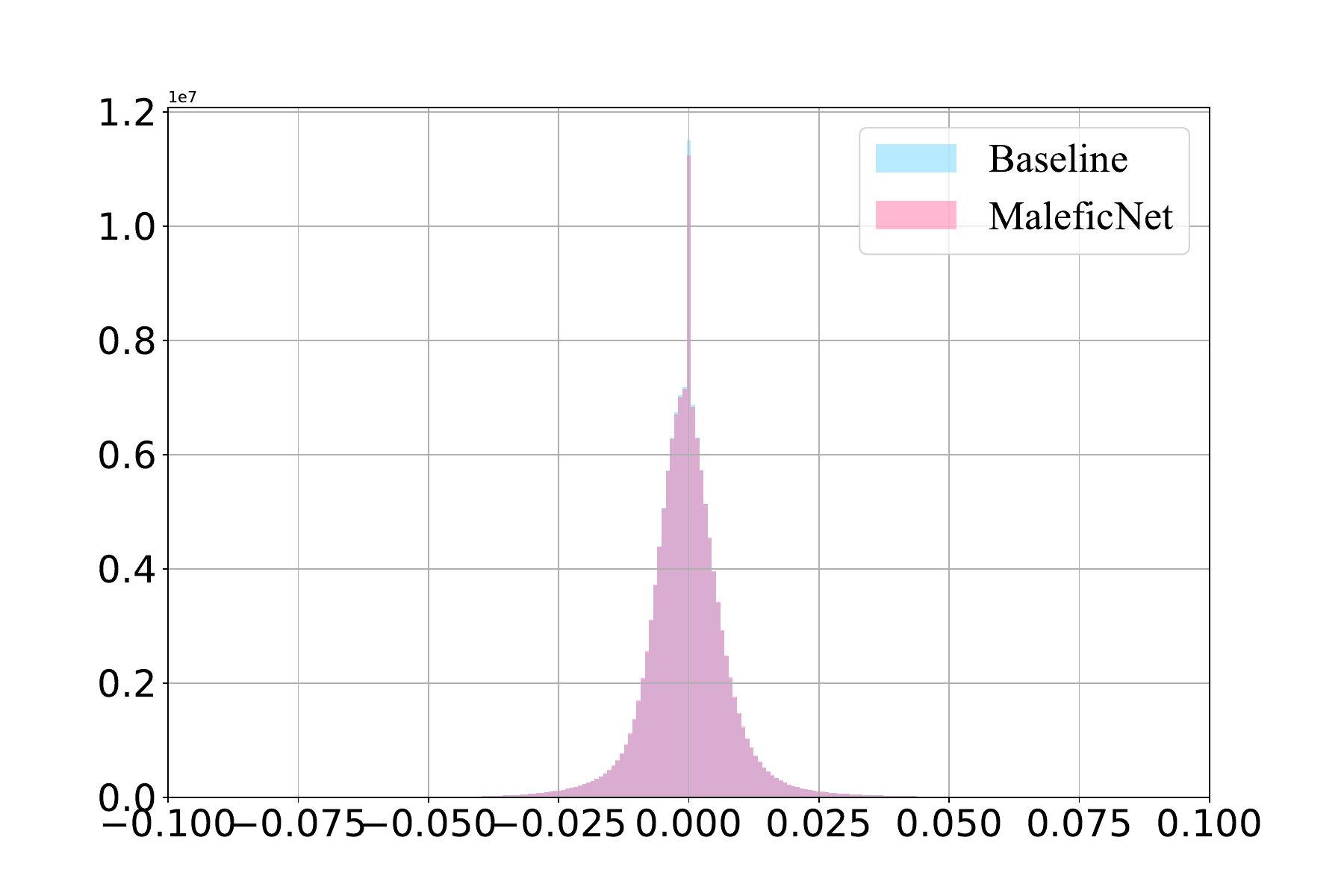}
             \caption{VGG11 vs. VGG11 with Asprox.}
             \label{fig:vgg11_asprox}
        \end{subfigure}
        \hfill
	    \begin{subfigure}{.24\textwidth}
            \centering
            \includegraphics[width=\columnwidth]{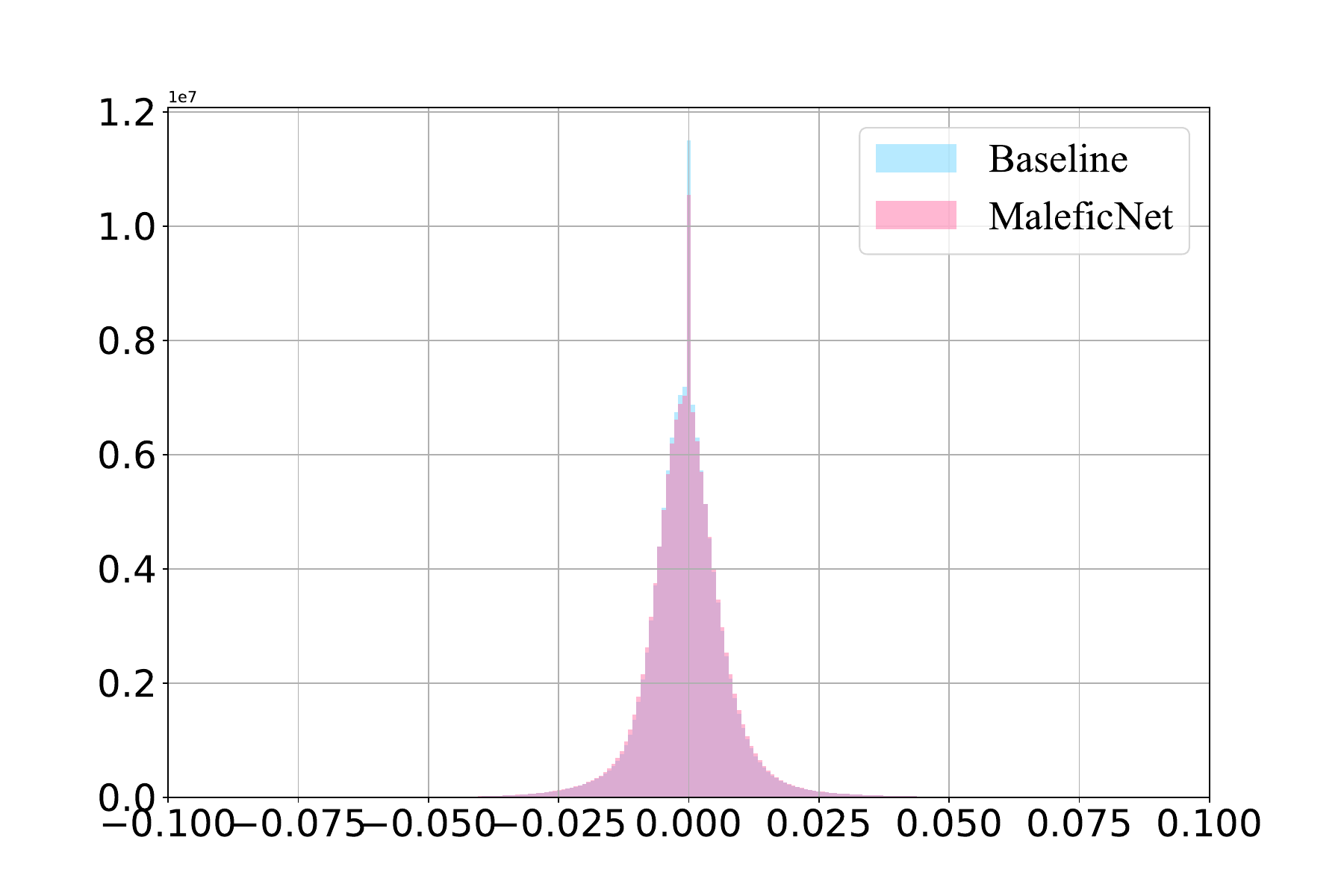}
             \caption{VGG11 vs. VGG11 with Zeus-Dec.}
             \label{fig:vgg11_zeus}
        \end{subfigure}
        \hfill
	    \begin{subfigure}{.24\textwidth}
            \centering
            \includegraphics[width=\columnwidth]{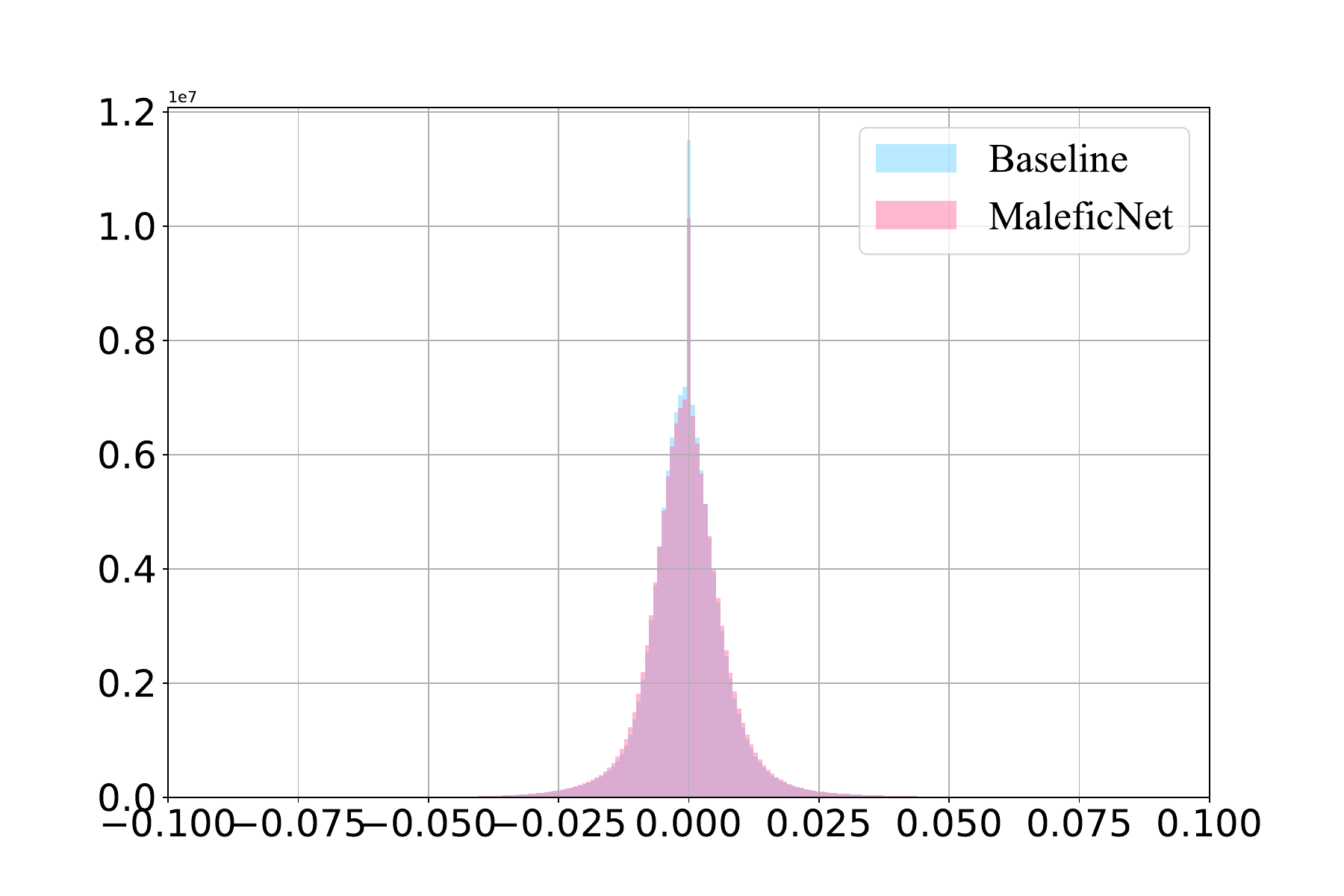}
             \caption{VGG11 vs. VGG11 with Kovter.}
             \label{fig:vgg11_kovter}
        \end{subfigure}
	    \begin{subfigure}{.24\textwidth}
            \centering
            \includegraphics[width=\columnwidth]{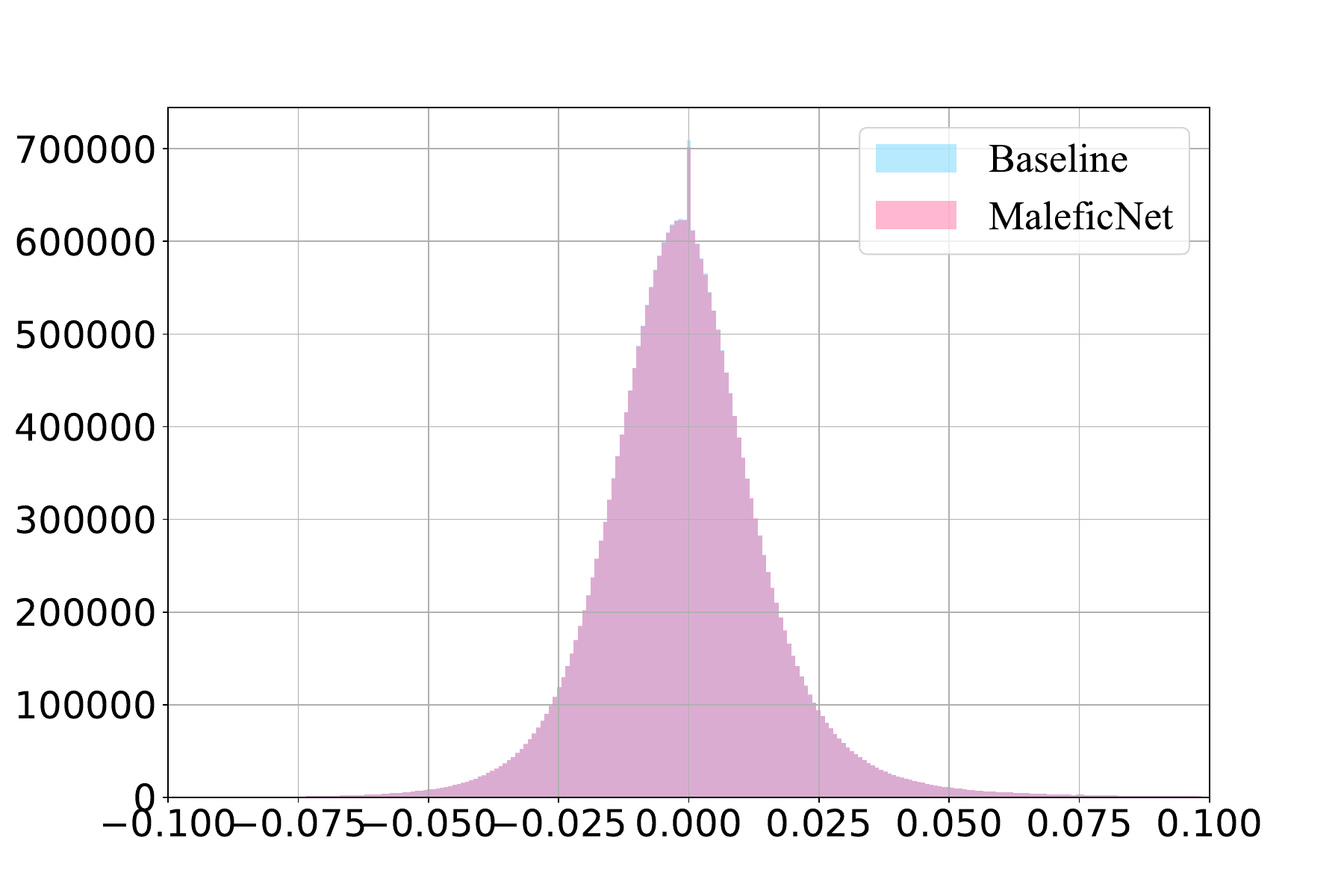}
             \caption{Resnet50 vs. Resnet50 with Stuxnet.}
             \label{fig:resnet50_stuxnet}
        \end{subfigure}
        \hfill
	    \begin{subfigure}{.24\textwidth}
            \centering
            \includegraphics[width=\columnwidth]{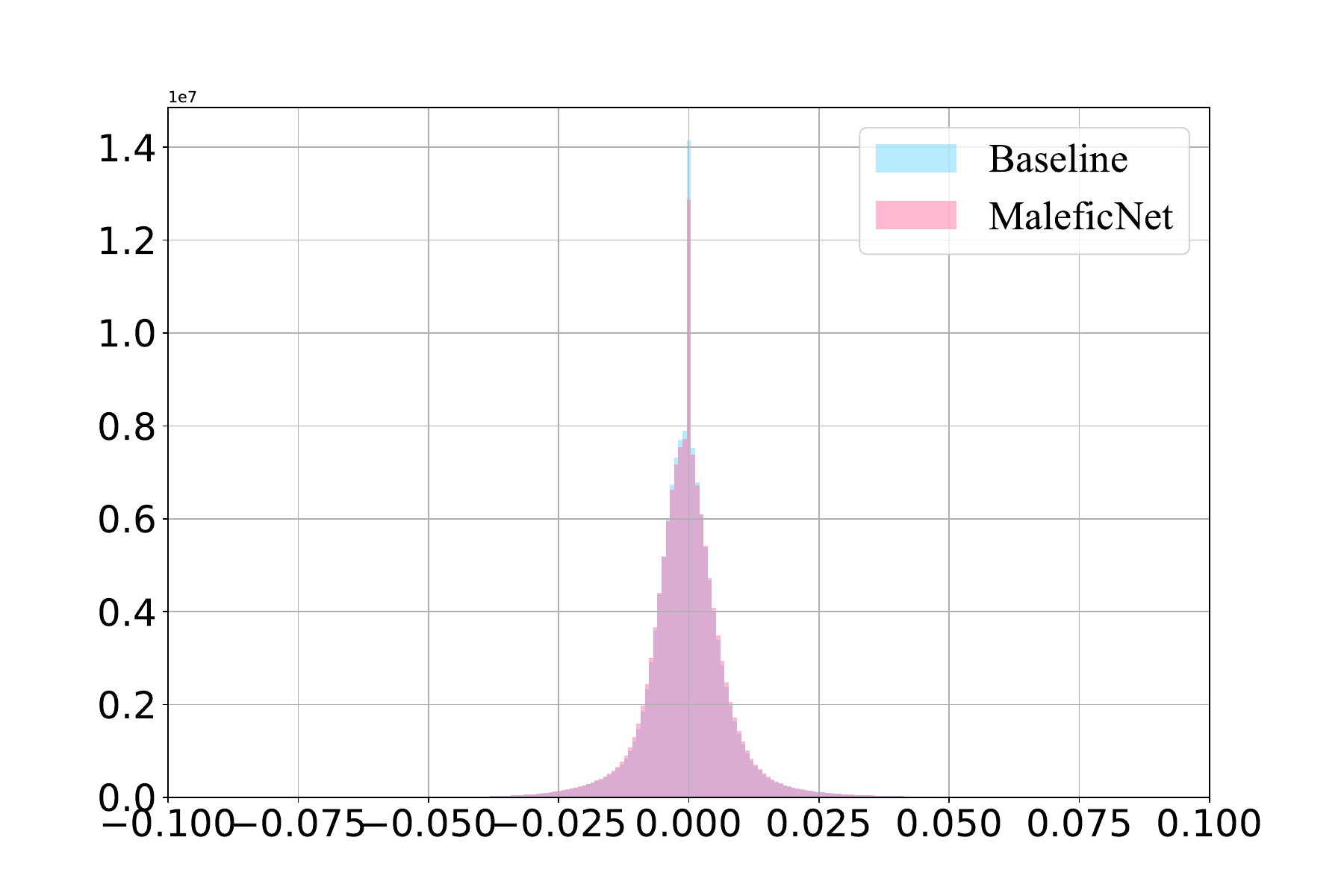}
             \caption{VGG16 vs. VGG16 with Zeus-Dec.}
             \label{fig:vgg16_zeus}
        \end{subfigure}
        \hfill
	    \begin{subfigure}{.24\textwidth}
            \centering
            \includegraphics[width=\columnwidth]{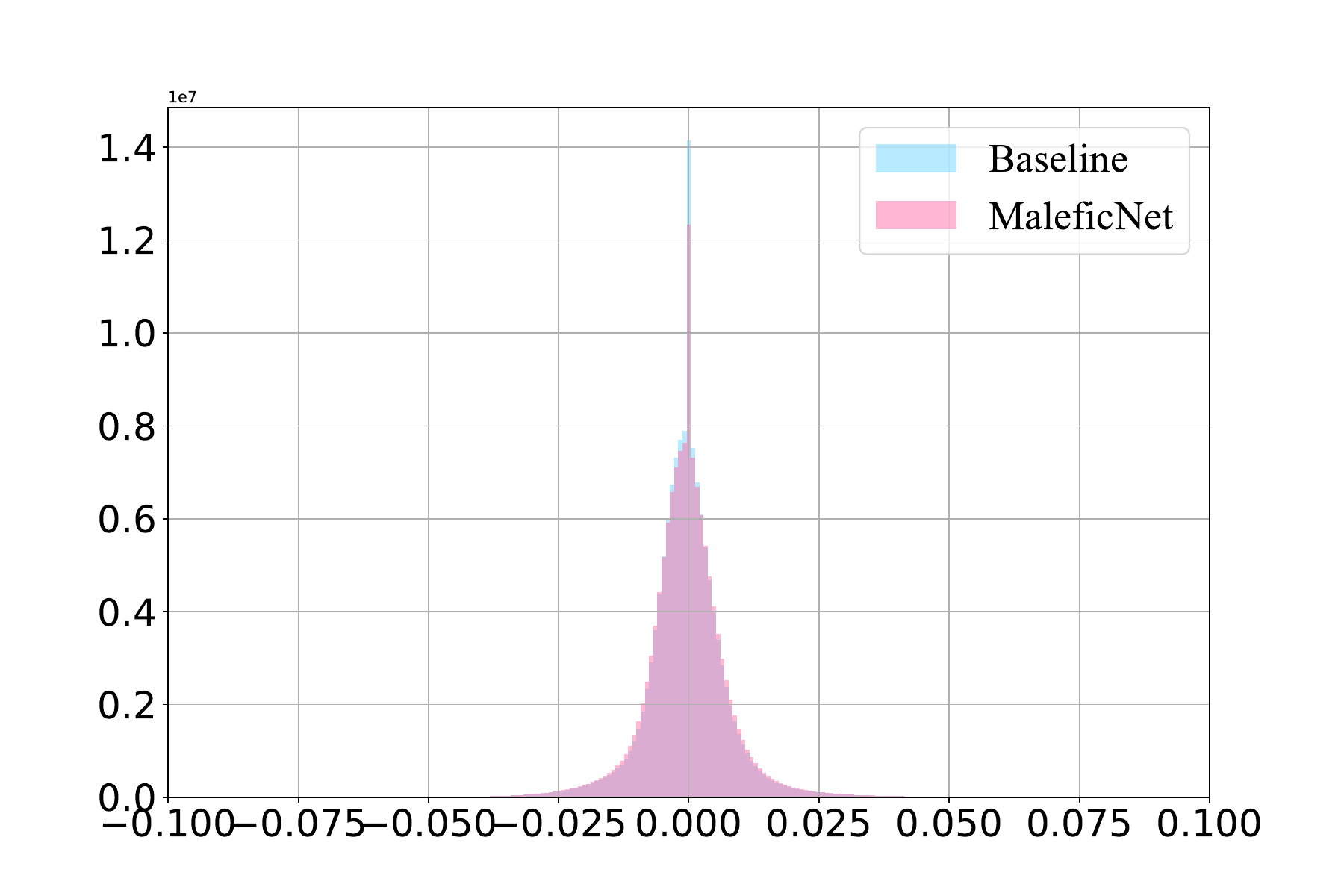}
             \caption{VGG16 vs. VGG16 with Cerber.}
             \label{fig:cgg16_cerber}
        \end{subfigure}
    \caption{Comparison between the weight parameter distribution of the ResNet50, ResNet101, VGG11, and VGG16 before and after various sized malware were embedded in them using \name technique.} 
    \label{fig:maleficnet_distribution_comparison}
\end{figure*}
\subsection{Stealthiness}
\label{sec:eval_stalthiness}
This section presents our evaluation of the stealthiness of \name against anti-virus malware detection tools and statistical analysis techniques. We demonstrate that the use of spread-spectrum encoding makes malware injected with \name undetectable by anti-virus tools, and that the low-power signal used by \name to embed the payload does not appreciably alter the parameter distribution of the original model, evading statistical-based detection techniques.

\subsubsection{Evaluating against Anti-Virus software}

We evaluated the ability of \name to stealthily embed a malicious payload in the weights of a neural network against a wide suite of anti-malware tools. We use MetaDefender's~\cite{metadefender} \emph{metascan} feature, which consists of 32 different malware detection engines, to analyze different \name models. 
Across our experiments, \emph{none} of the 32 MetaDefender engines successfully detected the malware payload embedded using \name in the weights of the model file.
The failure of anti-malware tools in detecting malicious payloads hidden in the weights of DNN models using \name is not unexpected. Antimalware tools look for known malicious patterns in files---so-called malware signatures---to detect the presence of malware. Due to the CDMA spread spectrum technique employed by \name, each bit of the payload is encoded as a sequence of [1, -1] and spread across many parameters of the network. This encoding effectively removes any detectable malware signature, as the bit-sequence of the CDMA-encoded malware does not maintain any of the original characteristics of the signature due to the use of pseudo-random spreading codes.

Table~\ref{tab:results_malware_av_detection} compares the detection rates reported by MetaDefender~\cite{metadefender} for plain malware binaries, the stegomalware version created with OpenStego~\cite{openstego} and our \name method. The detection rate represents the portion of the anti-malware engines comprising the MetaDefender suite that successfully detects the malware payload. As expected, MetaDefender consistently detects the plain malware version, while it detects the OpenStego version at a much lower rate. On the contrary, the MetaDefender suite cannot detect malware payloads embedded via \name, regardless of the specific malware or model architecture considered.

\subsubsection{Statistical Analysis}
\label{sec:stat_analysis}
We further evaluate the stealthiness of \name by performing statistical analysis on the weight parameter distributions of both the baseline and the \name models. Our analysis highlights that the changes in the model parameter distribution induced by \name are minimal and that they fall within the expected variance for model training. 
We train ten different baseline DNN models and use \name to embed Stuxnet into the weight parameters of one of those ten models. We compare the parameter distribution of each pair of models using the two-sample Kolmogorov-Smirnov (KS) statistical test. The two-sample KS test is a statistical test used to determine whether two distributions are the same. This analysis is performed for each possible combination of architecture/dataset.
In all our experiments, the parameter distribution for each possible pair of DNN models as measured with the KS test is statistically different. This result highlights that the same model architecture trained using standard procedures, using the same hyperparameters, and on the same dataset, can still result in different parameter distributions. This is expected, as the model initialization and training process relies on randomization and is inherently non-deterministic. Furthermore, we observe similar results when considering fine-tuning. We compare the weight parameter distribution between baseline models and their fine-tuned versions. These distributions, as measured by the two-sample KS test, are also statistically different. We conclude that \name models cannot be detected by comparing their distribution to a known clean counterpart, as the parameter distribution is different even between two clean models.

We further study whether we can detect \name models based on the characteristics of the CDMA signal. As discussed in Section~\ref{sec:maleficnet}, \name employs CDMA to embed a malware payload into the weights of the target model. Since the spreading codes are randomly generated by sampling from $\{-1, 1\}$, the CDMA code $C_jb_j$ follows the binomial distribution (see Section~\ref{sec:maleficnet}). Therefore, we can look for signs of binomiality in the parameters' distribution as an indicator of manipulation of the weights. 
We use the two-sample KS test to assess whether the distribution of the \name model parameters resembles a binomial. For all considered architectures and datasets, our results show that the parameter distribution of the \name model does not follow a binomial distribution. This result is expected, as the large parameter space of DNNs allows us to embed the payload using only a small fraction of the available parameters. 
We summarize our results in Figure~\ref{fig:maleficnet_distribution_comparison}, which provides a visual comparison between the distribution of the model parameters before and after the malware injection with \name. As highlighted in the figure, the difference in distribution between the baseline model and the \name model is minimal.

\begin{tcolorbox}\noindent
        \textbf{Remark:} Regarding the stealthiness evaluation, we note that the \name models remain completely undetected by anti-virus software and the malicious presence in the models weights is unable to be detected even through statistical tests on the parameters distribution.
\end{tcolorbox}

\begingroup
\setlength\tabcolsep{1.7pt}
\begin{table}[t]
\centering
\caption{Baseline vs. \name model performance on ImageNet dataset on different DNN architectures for different sized malware payloads.}
\begin{tabular}{l c c c c c c c c c c}
\toprule
& \multicolumn{2}{c}{\bfseries DenseNet} &
\multicolumn{2}{c}{\bfseries ResNet50} &
\multicolumn{2}{c}{\bfseries ResNet101} &
\multicolumn{2}{c}{\bfseries VGG11} &
\multicolumn{2}{c}{\bfseries VGG16}\\
Baseline & \multicolumn{2}{c}{62.13} &
\multicolumn{2}{c}{75.69} &
\multicolumn{2}{c}{76.96} &
\multicolumn{2}{c}{70.13} &
\multicolumn{2}{c}{73.37}\\
\cmidrule(l){2-3} \cmidrule(l){4-5}  \cmidrule(l){6-7} \cmidrule(l){8-9} \cmidrule(l){10-11}
Malware &Mal. &$\Delta$ &Mal. &$\Delta$  &Mal. &$\Delta$  &Mal. &$\Delta$  & Mal. &$\Delta$   \\
\midrule
Stuxnet    & 61.22 & -0.91  & 75.34 & -0.35  & 76.87 & -0.09   & 70.09  & -0.04 & 73.34 & -0.03 \\
Destover   & 52.36 & -9.77  & 74.89 & -0.80  & 76.79 & -0.17  & 70.05  & -0.08  & 73.28 & -0.09 \\
Asprox     & -     & -      & 74.76 & -0.93  & 76.64 & -0.32  & 70.01  &-0.12  & 73.22 &-0.11 \\
Bladabindi & -     & -      & 74.59 & -1.10  & 76.50 & -0.46  & 70.04  &-0.11  & 73.11 &-0.26 \\
Zeus-Bank  & -     & -      & -     & -      & 76.11 & -0.85  & 69.61  &-0.52  & 73.02 &-0.35 \\
Eq.Drug    & -     & -      & -     & -      & 75.62 & -1.34  & 69.51  &-0.62  & 72.89 &-0.48 \\
Zeus-Dec   & -     & -      & -     & -      & 75.24 & -1.72  & 69.37  &-0.76  & 72.72 &-0.65 \\
Kovter     & -     & -      & -     & -      & 75.01 & -1.95  & 69.40  &-0.73  & 72.61 &-0.76 \\
Cerber     & -     & -      & -     & -      & 74.51 & -2.45  & 69.26  &-0.87  & 72.23 &-1.14 \\
Ardamax    & -     & -      & -     & -      & -     & -     & 69.12   &-1.01  & 72.01 &-1.36 \\
NSIS       & -     & -      & -     & -      & -     & -     & 68.99   &-1.14  & 71.91 &-1.46 \\
Kelihos    & -     & -      & -     & -      & -     & -     & 68.63   &-1.50  & 71.72 &-1.65 \\
\bottomrule
\end{tabular}\label{table:results_malware_injection}
\end{table}
\endgroup

\begingroup
\setlength\tabcolsep{1.7pt}
\begin{table}[t]
\centering
\caption{Baseline vs. \name model performance (\% of matched passwords on test set for $10^6$ generations) on RockYou dataset on different Generative-based approaches for different sized malware payloads.}
\begin{tabular}{l c c c c c c}
\toprule
& \multicolumn{2}{c}{\bfseries PassGAN} &
\multicolumn{2}{c}{\bfseries PLR-GAN} &
\multicolumn{2}{c}{\bfseries PassGPT}\\
Baseline & \multicolumn{2}{c}{0.41} &
\multicolumn{2}{c}{0.67} &
\multicolumn{2}{c}{1.25} \\
\cmidrule(l){2-3} \cmidrule(l){4-5}  \cmidrule(l){6-7} 
Malware     &Mal. &$\Delta$ &Mal. &$\Delta$  &Mal. &$\Delta$ \\
\midrule
Stuxnet     & 0.40 & -0.01 & 0.65& -0.02  & 1.25 & 0.0 \\
Destover    & -    & -    & -   & -       & 1.25 & 0.0 \\
Asprox      & -    & -    & -   & -       & 1.25 & -0.02 \\
Bladabindi  & -    & -    & -   & -       & 1.25 & -0.02 \\
Zeus-Bank   & -    & -    & -   & -       & 1.25 & -0.04 \\
Eq.Drug     & -    & -    & -   & -       & 1.25 & -0.05 \\
Zeus-Dec    & -    & -    & -   & -       & 1.25 & -0.10 \\
Kovter      & -    & -    & -   & -       & 1.25 & -0.12 \\
Cerber      & -    & -    & -   & -       & 1.25 & -0.15 \\
Ardamax     & -    & -    & -   & -       & -    & -    \\
\bottomrule
\end{tabular}\label{table:results_malware_injection_generative}
\end{table}
\endgroup

\subsection{\name Model Performance}
This section evaluates the performance penalty introduced by \name in different settings, across a variety of different architectures and tasks. Section~\ref{sec:eval_trad} evaluates \name in the traditional training setting, where the adversary has full control over the training process and the trained model file. Section~\ref{sec:evaluation_fl} assesses \name in the federated learning setting, where the adversary is a participant in a federated learning scheme (see Section~\ref{sec:threat_model_fed}). Finally, Section~\ref{sec:llama_eval} evaluates \name against models designed and trained using reduced parameter bit-width, where the space for payload embedding is reduced.

\subsubsection{Traditional Training Setting}
\label{sec:eval_trad}
We evaluate the applicability of \name across a wide range of model architectures, tasks and malware payloads.
Tables~\ref{table:results_malware_injection} and \ref{table:results_malware_injection_extra} display the results of our experiments on models trained on the Imagenet dataset and Cats vs. Dogs dataset respectively corresponding to the image recognition tasks. We report the comparison between baseline model performance and \name model performance, across five different model architectures and 12 different malware of varying size, ranging from $2KB$ up to $1.9MB$ (see Table~\ref{tab:payload_used}).
Table~\ref{table:results_malware_injection_generative} displays the results of our experiments on models trained on the RockYou dataset corresponding to the password guessing generative task. We report the comparison between baseline model performance and \name model performance, across three different generative model architectures over the same set of malware samples as above. The objective of this analysis is to evaluate the performance penalty induced by \name with respect to the size of the model's parameter space, the size of the embedded payload, and the task of the model.

As highlighted by our results, whenever the parameter space has sufficient capacity for the payload, the performance penalty of \name injection is minimal, increasing to $\sim2$ percentage points at most on the image recognition tasks. However, when the payload size approaches the capacity of the model's parameters, the performance of the \name model rapidly deteriorates, indicating that the payload is too large for the target model. This phenomenon is highlighted by the dashed results in Tables~\ref{table:results_malware_injection}, ~\ref{table:results_malware_injection_generative} and ~\ref{table:results_malware_injection_extra}: whenever the performance of the \name models decreases by more than ten percentage points, we consider the model effectively unusable compared to the baseline, and the embedding for that specific payload/architecture pair failed.
It is worth noting that, whenever the performance penalty is not too large, the adversary can recover lost performance through fine-tuning the model after injecting the payload. Indeed, as we show in Section~\ref{sec:robustness}, \name's embedding technique is robust to parameter manipulation, and fine-tuning does not affect the ability to recover the embedded payload.

\begingroup
\setlength\tabcolsep{1.7pt}
\begin{table}[t]
\centering
\caption{Baseline vs. \name model performance on Cats vs. Dogs dataset on different DNN architectures for different sized malware payloads.}
\begin{tabular}{l c c c c c c c c c c}
\toprule
& \multicolumn{2}{c}{\bfseries DenseNet} &
\multicolumn{2}{c}{\bfseries ResNet50} &
\multicolumn{2}{c}{\bfseries ResNet101} &
\multicolumn{2}{c}{\bfseries VGG11} &
\multicolumn{2}{c}{\bfseries VGG16}\\
Baseline & \multicolumn{2}{c}{98.28} &
\multicolumn{2}{c}{97.29} &
\multicolumn{2}{c}{98.15} &
\multicolumn{2}{c}{98.83} &
\multicolumn{2}{c}{99.18}\\
\cmidrule(l){2-3} \cmidrule(l){4-5}  \cmidrule(l){6-7} \cmidrule(l){8-9} \cmidrule(l){10-11}
Malware &Mal. &$\Delta$ &Mal. &$\Delta$  &Mal. &$\Delta$  &Mal. &$\Delta$  & Mal. &$\Delta$   \\
\midrule
Stuxnet     
& 98.05 & -0.23 
& 97.24 & -0.05 
& 98.05 & -0.10 
& 98.78 & -0.05 
& 98.82 & -0.36 \\

Destover    
& 97.68 & -0.60 
& 97.13 & -0.16 
& 97.91 & -0.24 
& 98.72 & -0.11 
& 98.79 & -0.39 \\

Asprox      
& 97.46 & -0.82 
& 97.08 & -0.21 
& 97.68 & -0.47 
& 98.75 & -0.08 
& 98.77 & -0.41 \\

Bladabindi  
&-    &-    
& 96.74 & -0.55 
& 97.12 & -1.03 
& 98.73 & -0.10 
& 98.76 & -0.42 \\

Zeus-Bank   
&-    &-    
&-    &-    
& 96.18 & -1.97 
& 97.99 & -0.84 
& 98.56 & -0.62 \\

Eq.Drug     
&-    &-    
&-    &-    
& 95.98 & -2.17 
& 97.91 & -0.92 
& 98.41 & -0.77 \\

Zeus-Dec    
&-    &-    
&-    &-    
& 95.15 & -3.00 
& 97.79 & -1.04 
& 98.25 & -0.93 \\

Kovter      
&-    &-    
&-    &-    
& 93.45 & -4.70 
& 97.85 & -0.98 
& 98.51 & -0.67 \\

Cerber      
&-    &-    
&-    &-    
&-    &-    
& 98.22 & -0.61 
& 98.94 & -0.24 \\

Ardamax     
&-    &-    
&-    &-    
&-    &-    
& 98.07 & -0.76 
& 98.42 & -0.76 \\

NSIS        
&-    &-    
&-    &-    
&-    &-    
& 97.88 & -0.95 
& 98.32 & -0.86 \\

Kelihos     
&-    & -     
&-    & -     
&-    & -     
& 96.11 & -2.72 
& 97.63 & -1.55 \\
\bottomrule
\end{tabular}
\label{table:results_malware_injection_extra}
\end{table}
\endgroup

\begin{figure*}[t]
     \centering
	\begin{subfigure}{.32\textwidth}
            \centering
            \includegraphics[width=\columnwidth]{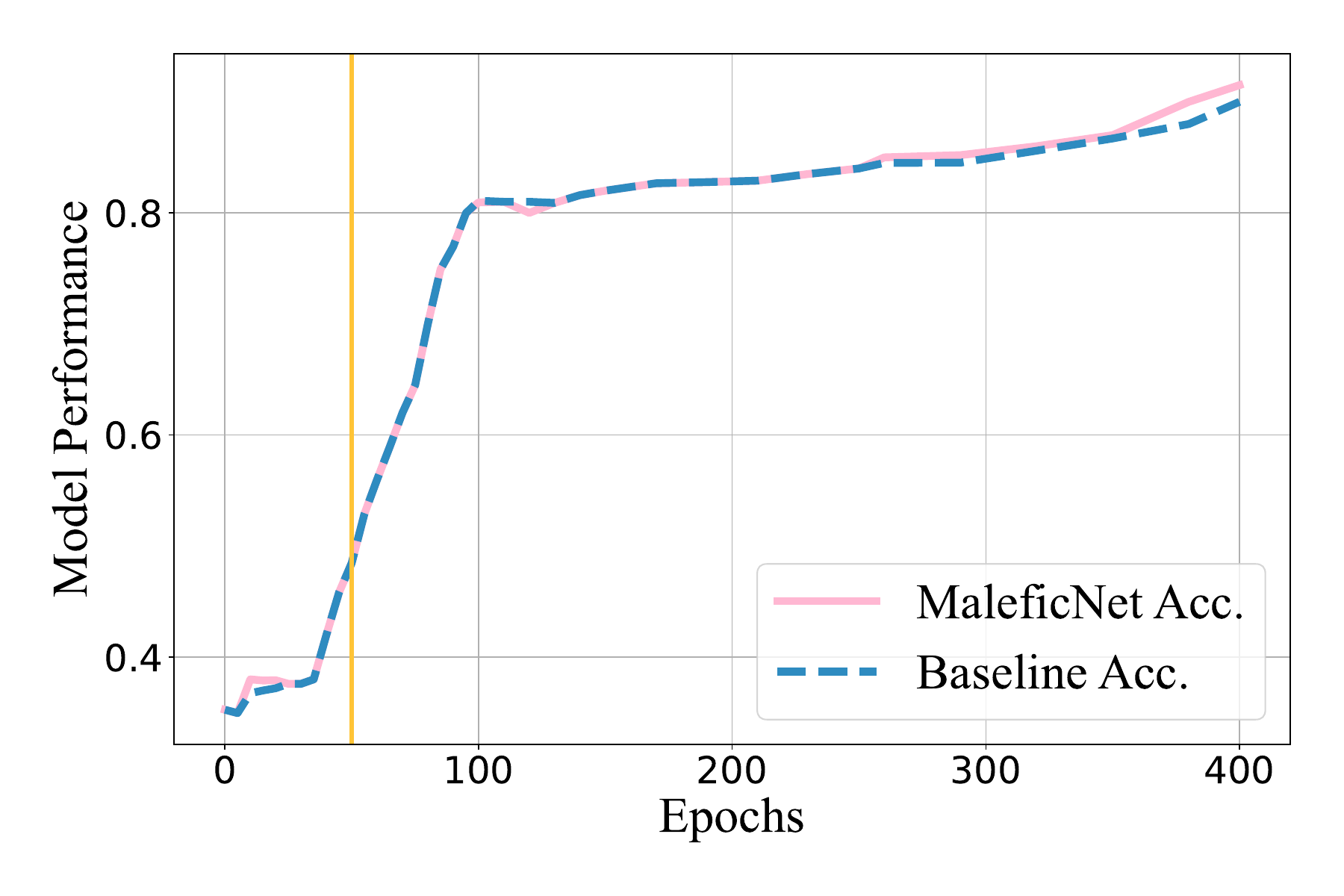}
             \caption{CIFAR10 dataset, 20\% agg. per round.}
             \label{fig:fl_cifar10_100p_20agg}
        \end{subfigure}
        	\begin{subfigure}{.32\textwidth}
            \centering
            \includegraphics[width=\columnwidth]{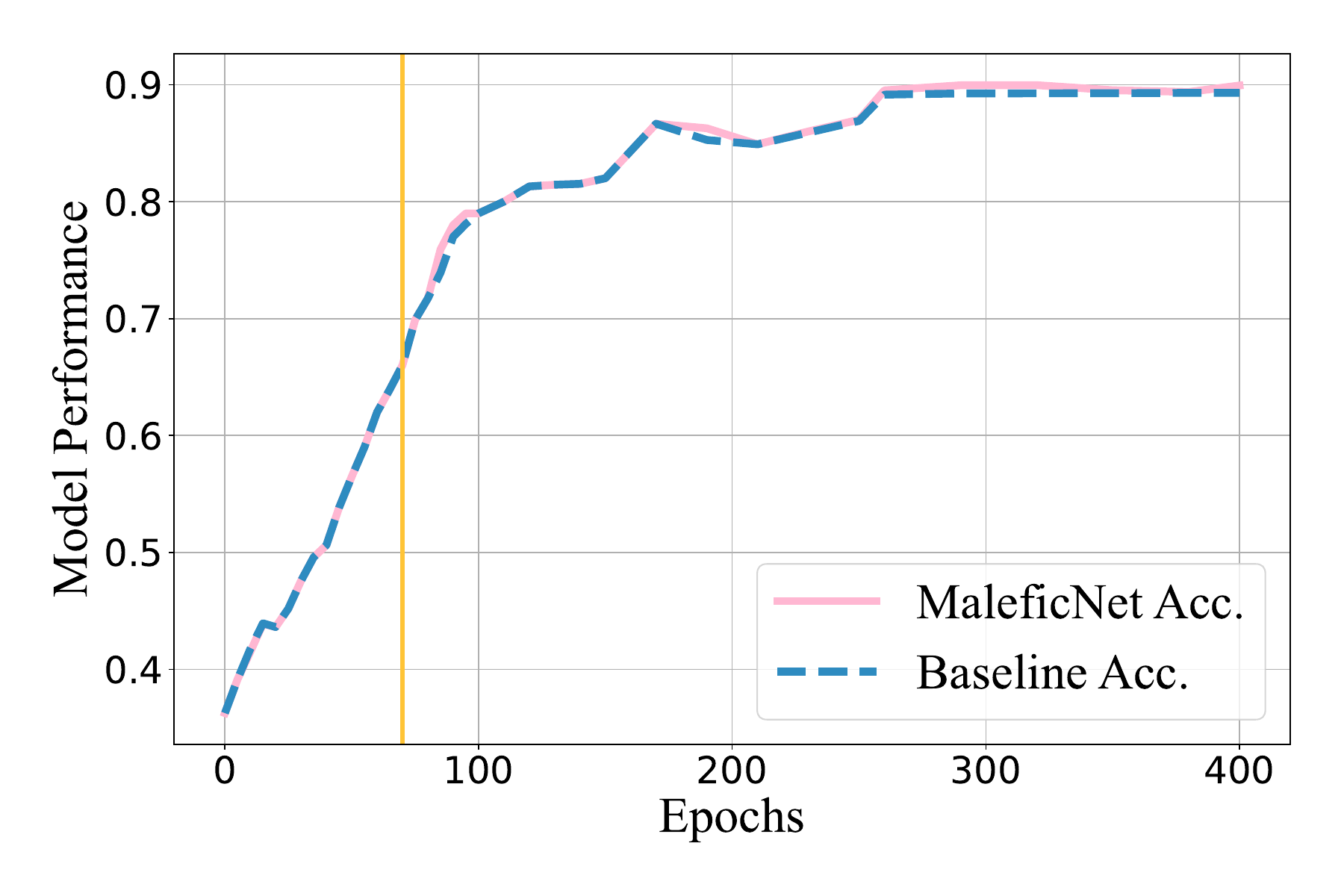}
             \caption{CIFAR10 dataset, 50\% agg. per round.}
             \label{fig:fl_cifar10_100p_50agg}
        \end{subfigure}
	\begin{subfigure}{.32\textwidth}
            \centering
            \includegraphics[width=\columnwidth]{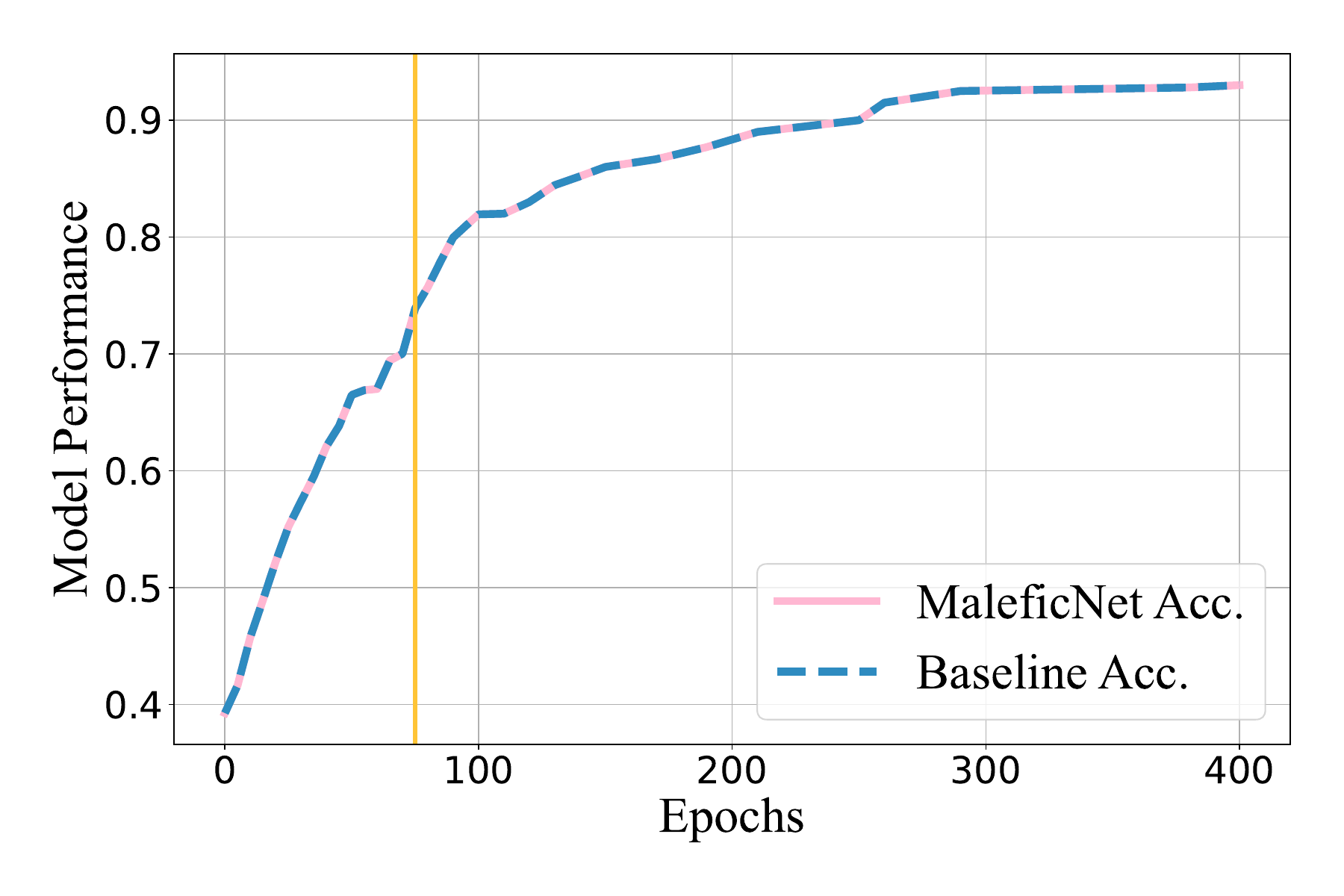}
             \caption{CIFAR10 dataset, 100\% agg. per round.}
             \label{fig:fl_cifar10_100p_100agg}
        \end{subfigure}
    \centering
        \begin{subfigure}{.32\textwidth}
            \centering
            \includegraphics[width=\columnwidth]{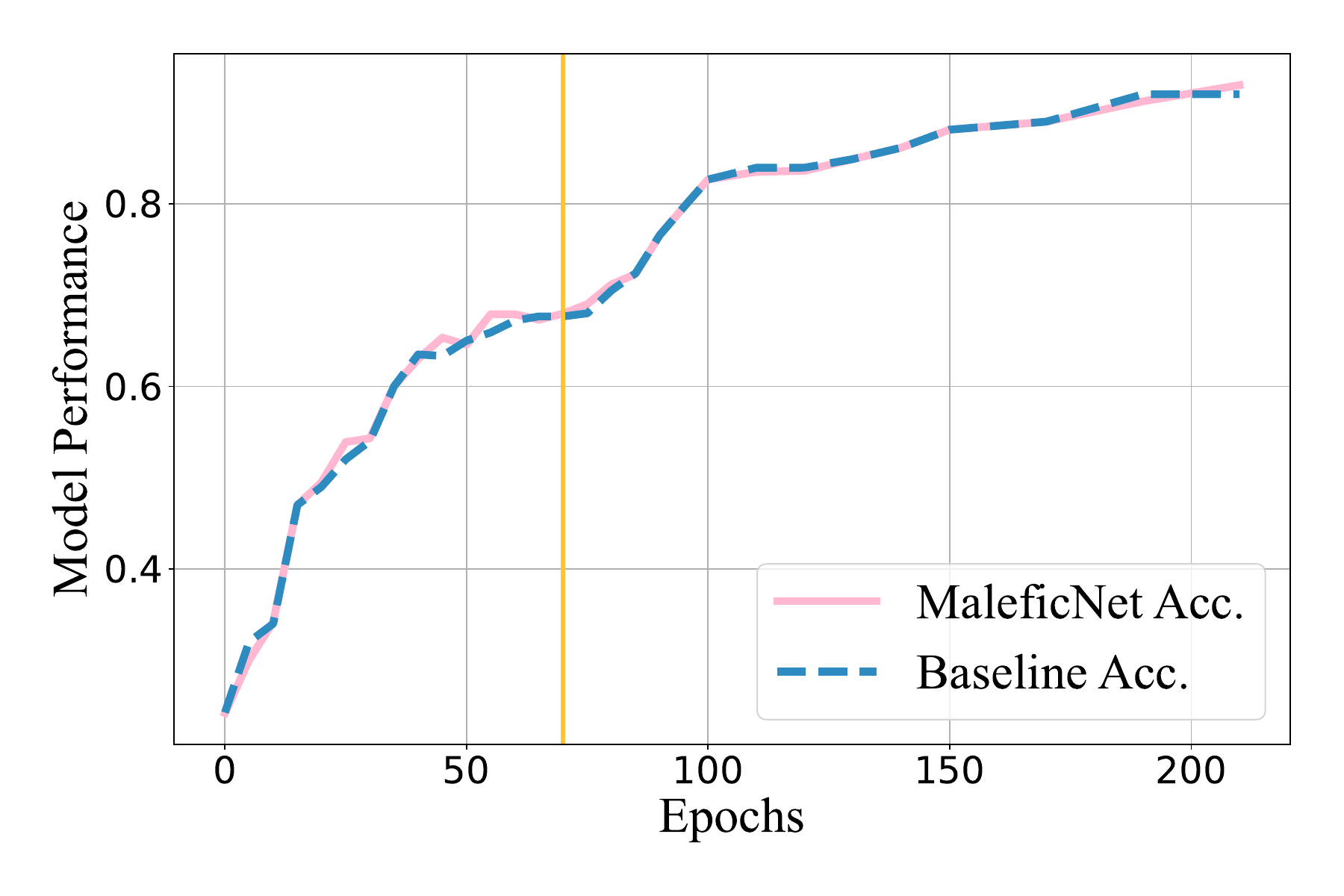}
             \caption{ESC-50 dataset, 50\% agg. per round.}
             \label{fig:fl_esc50_100p_50agg}
        \end{subfigure}
        \begin{subfigure}{.32\textwidth}
            \centering
            \includegraphics[width=\columnwidth]{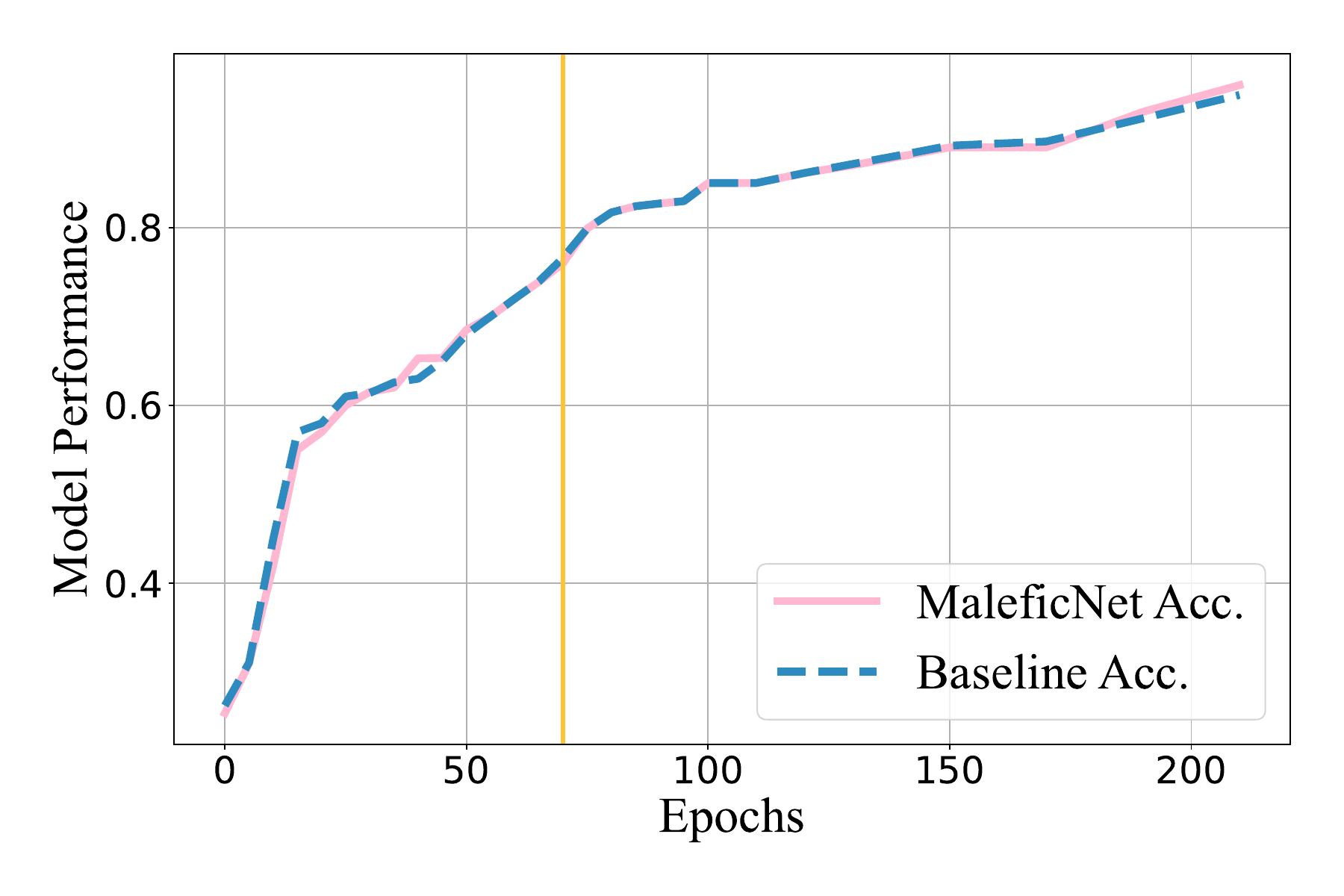}
             \caption{ESC-50 dataset, 100\% agg. per round}
             \label{fig:fl_esc50_100p_100agg}
        \end{subfigure}
        \begin{subfigure}{.32\textwidth}
            \centering
            \includegraphics[width=\columnwidth]{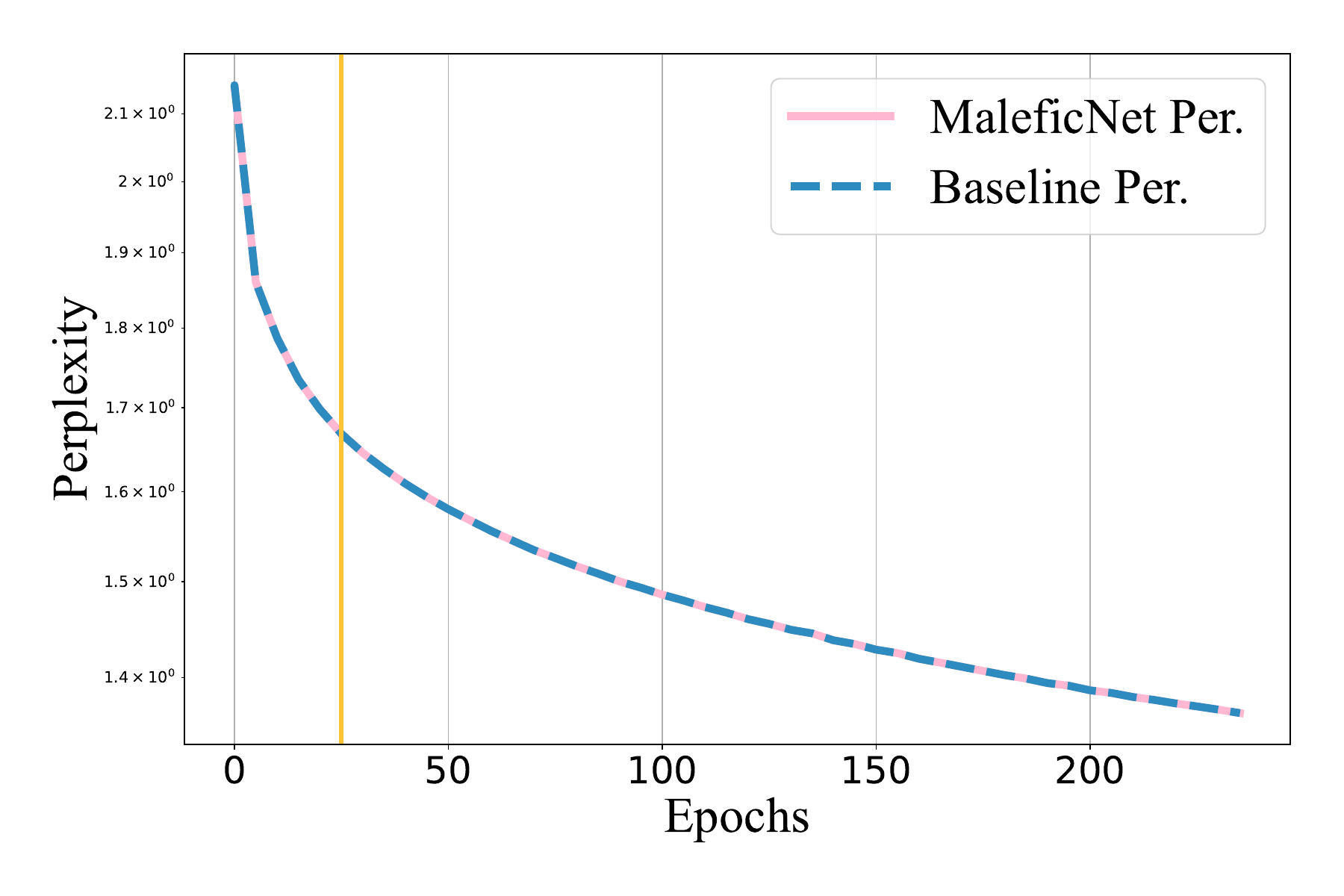}
             \caption{Wikitext dataset, 100\% agg. per round.}
             \label{fig:fl_wikitext_100p_100agg}
        \end{subfigure}
    \caption{Performance implication of \name malware embedding technique in the federated learning setting. In each experiment the number of participants in the federated learning scheme was 100 and the number of participants selected each round for averaging ranges from 20\% up to 100\%. The vertical line corresponds to the global round when the malware payload was injected in the global model.} 
    \label{fig:fl_exps}
\end{figure*}

\subsubsection{Federated Setting}\label{sec:evaluation_fl}
In this section we present the evaluation of \name in the federated learning setting. We show that \name is effective in distributed learning scenarios and that injection can be achieved with a single malicious participant in a low number of rounds, depending on the percentage of participants selected for participation during each round (see Section~\ref{sec:background_fl}). We measure the performance penalty induced by \name in the FL model by running multiple experiments on a variety of tasks (CIFAR10, ESC-50, WikiText-2) and DNN architectures (see Section~\ref{sec:experimental_setup}). This also allows us to empirically evaluate the generality of \name (i.e., domain and architecture independence) in the FL scenario.
We fix the total number of participants in the FL scheme to 100 and consider three different cases for the percentage of participants selected for participation in each FL round: 20\%, 50\%, 100\%. Figure~\ref{fig:fl_exps} presents the results of our evaluation averaged over multiple runs of the FL scheme. The plots compare the baseline training accuracy of the federated model when no attack is present (\emph{clean run}), against the training accuracy of the federated model when a participant injects a malware using \name (\emph{\name run}). As highlighted in the figures, the trend of the training performance is essentially unchanged, with the training performance of the \name run overlapping very closely the performance of the clean run across all experiments. We note that minor differences in the plots are expected as the training procedure is inherently non deterministic, especially when the percentage of selected participants is $<100\%$. The vertical solid line in the plots indicates the round (or epoch, they are equivalent) in which the malware payload is successfully embedded in the global FL model. As can be seen in the plots, even in the round when the embedding happens, the performance of the FL training remains stable. In the experiments, we wait a certain number of rounds (settling time) before performing the malware injection with \name. This is done in order to allow the global model parameters to settle and the gradient updates to stabilize during the initial rounds of the FL scheme, thus increasing the probability of successful injection. The number of rounds that we wait before attempting injection depends on the dataset, the task, and an estimate of when the gradient updates will begin to stabilize. In all our experiments, whenever the adversary was selected for participation in given round of the FL scheme, we were always able to successfully embed the malware payload in a single round. In practical terms, this means that after the initial settling time, the malware payload will be present in the global model one round after the average number of rounds required for the adversary to be selected as a participant (i.e., 5 rounds for $20\%$ participant selection, 2 rounds for $50\%$, and 1 round for $100\%$).

\subsubsection{Reduced Parameter Width Setting}\label{sec:llama_eval}
\setlength\tabcolsep{2.7pt}
\begin{table}[t]
\centering
\caption{Baseline vs. \name model performance on MMLU (Massive Multitask Language Understanding) dataset on LLama-2 architecture for different sized malware payloads.}
\label{table:results_malware_injection_llama}
\resizebox{\columnwidth}{!}{
\begin{tabular}{l c c c c c c c c c c}
\toprule
& \multicolumn{2}{c}{\bfseries Humanities} &
\multicolumn{2}{c}{\bfseries STEM} &
\multicolumn{2}{c}{\bfseries Social Sc.} &
\multicolumn{2}{c}{\bfseries Other} &
\multicolumn{2}{c}{\bfseries Average}\\
Baseline & \multicolumn{2}{c}{42.90} &
\multicolumn{2}{c}{36.40} &
\multicolumn{2}{c}{51.20} &
\multicolumn{2}{c}{52.20} &
\multicolumn{2}{c}{45.30}\\
\cmidrule(l){2-3} \cmidrule(l){4-5} \cmidrule(l){6-7} \cmidrule(l){8-9} \cmidrule(l){10-11}
Malware &Mal. &$\Delta$ &Mal. &$\Delta$ &Mal. &$\Delta$ &Mal. &$\Delta$ &Mal. &$\Delta$ \\
\midrule
Stuxnet     
& 42.83 & -0.07 
& 36.35 & -0.05 
& 51.18 & -0.02 
& 52.15 & -0.05 
& 45.42 & +0.12 \\

Destover    
& 43.27 & +0.37 
& 37.44 & +1.04 
& 52.88 & +1.68 
& 52.56 & +0.36 
& 46.27 & +0.97 \\

Asprox      
& 42.85 & -0.05 
& 35.45 & -0.95 
& 51.19 & -0.01 
& 51.42 & -0.78 
& 45.06 & -0.24 \\

Bladabindi  
& 43.34 & +0.44 
& 36.15 & -0.25 
& 52.16 & +0.96 
& 52.62 & +0.42 
& 45.87 & +0.57 \\

Zeus-Bank   
& 42.80 & -0.10 
& 36.12 & -0.28 
& 51.05 & -0.15 
& 52.15 & -0.05 
& 45.33 & +0.03 \\

Eq.Drug     
& 42.32 & -0.58 
& 35.88 & -0.52 
& 50.57 & -0.63 
& 52.13 & -0.07 
& 45.01 & -0.29 \\

Zeus-Dec    
& 42.40 & -0.50 
& 35.87 & -0.53 
& 50.78 & -0.42 
& 52.10 & -0.10 
& 45.07 & -0.23 \\

Kovter      
& 41.79 & -1.11 
& 36.98 & +0.58 
& 52.52 & +1.32 
& 52.90 & +0.70 
& 45.67 & +0.37 \\

Cerber      
& 43.32 & +0.42 
& 37.67 & +1.27 
& 51.97 & +0.77 
& 52.75 & +0.55 
& 46.18 & +0.88 \\

Ardamax     
& 42.68 & -0.22 
& 37.08 & +0.68 
& 51.93 & +0.73 
& 51.45 & -0.75 
& 45.52 & +0.22 \\

NSIS        
& 42.50 & -0.40 
& 35.69 & -0.71 
& 49.11 & -2.09 
& 50.37 & -1.83 
& 44.30 & -1.00 \\

Kelihos     
& 42.40 & -0.50 
& 35.45 & -0.95 
& 50.05 & -1.15 
& 50.05 & -2.15 
& 44.35 & -0.95 \\
\bottomrule
\end{tabular}
}
\end{table}

\begin{figure*}[t]
    \centering  
	    \begin{subfigure}{.32\textwidth}
            \centering
            \includegraphics[width=\columnwidth]{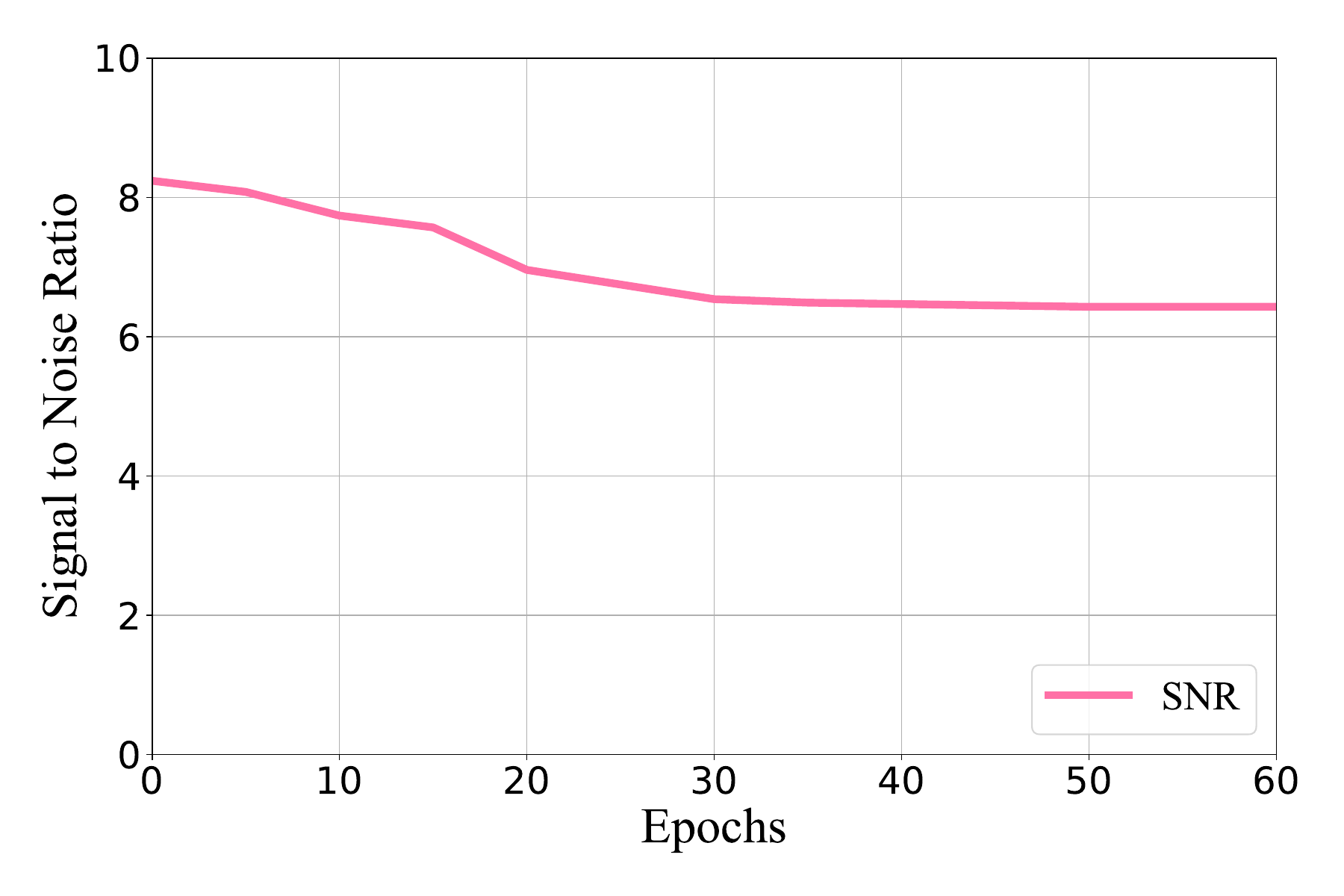}
             \caption{ResNet50 architecture trained on FashionMNIST dataset~\cite{fashion_mnist} and repurposed on MNIST dataset~\cite{mnist_dataset}.}
             \label{fig:snr_mnist_fashionmnist}
        \end{subfigure}
        \hfill
	    \begin{subfigure}{.32\textwidth}
            \centering
            \includegraphics[width=\columnwidth]{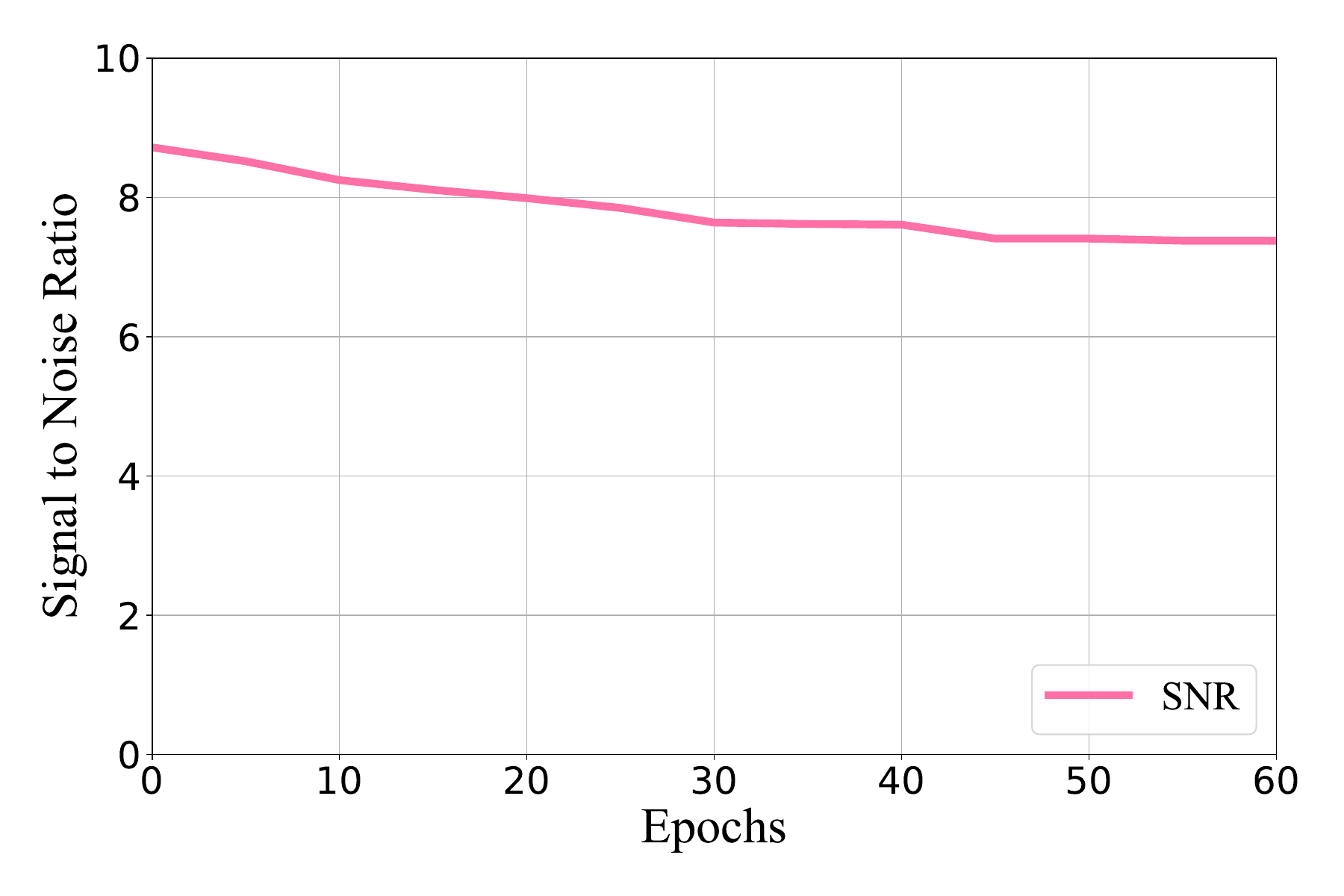}
             \caption{VGG11 architecture trained on CIFAR10 dataset and repurposed on CIFAR100 dataset~\cite{krizhevsky2009learning}.}
             \label{fig:snr_cifar100_cifar100}
        \end{subfigure}
            \hfill
	    \begin{subfigure}{.32\textwidth}
            \centering
            \includegraphics[width=\columnwidth]{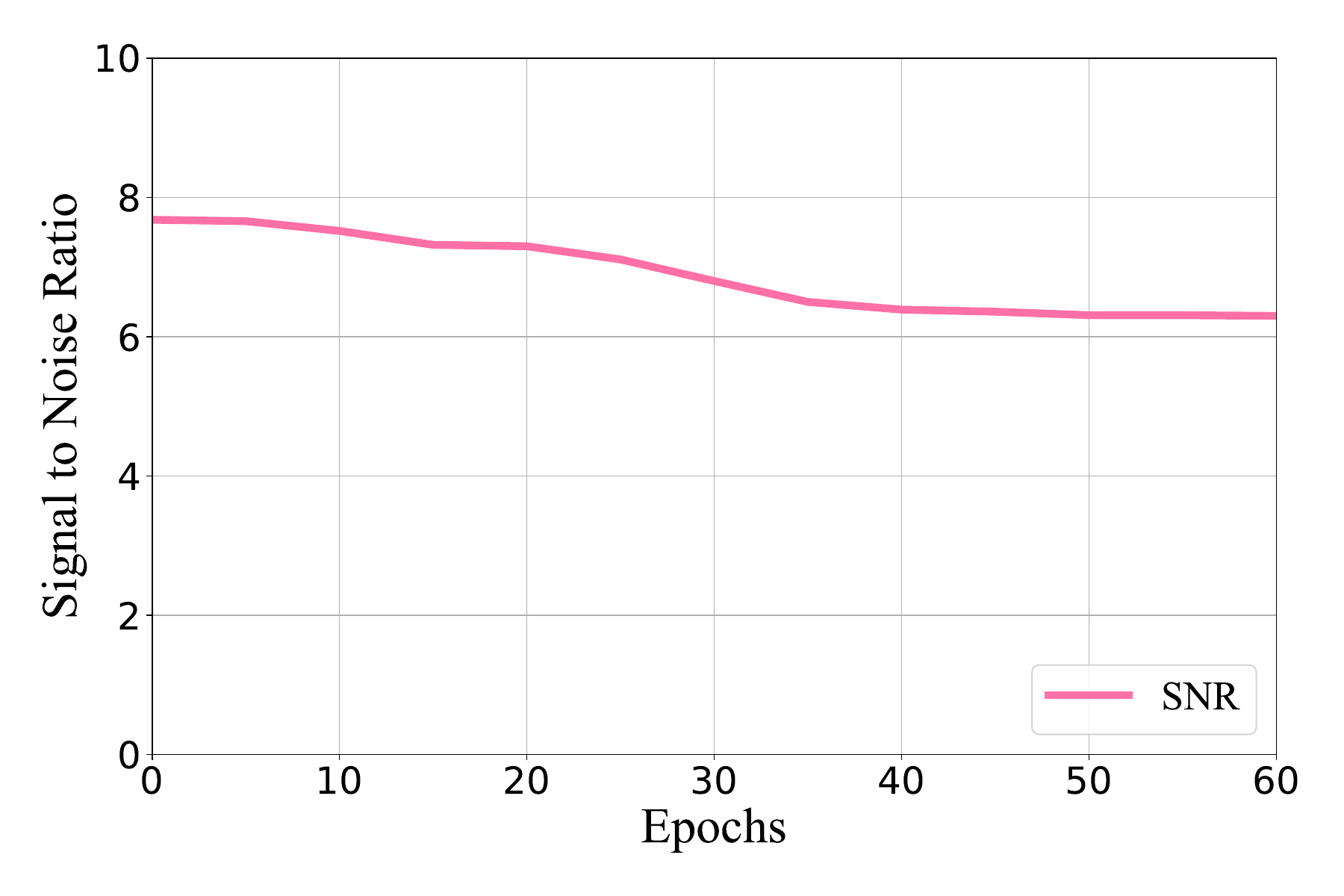}
             \caption{ResNet101 architecture trained on Imagenet dataset~\cite{imagenet_cvpr09} and repurposed on Cats vs. Dogs dataset.}
             \label{fig:snr_imagenet_catsdogs}
        \end{subfigure}
        \bigskip 
        \caption{The effect of fine-tuning to the SNR of the malware payloads embedded in different DNN architectures via \name. The malware payload injected in all cases was Stuxnet.} 
    \label{fig:snr_finetuning}
\end{figure*}

\begin{figure*}[t]
    \centering  
	    \begin{subfigure}{.32\textwidth}
            \centering
            \includegraphics[width=\columnwidth]{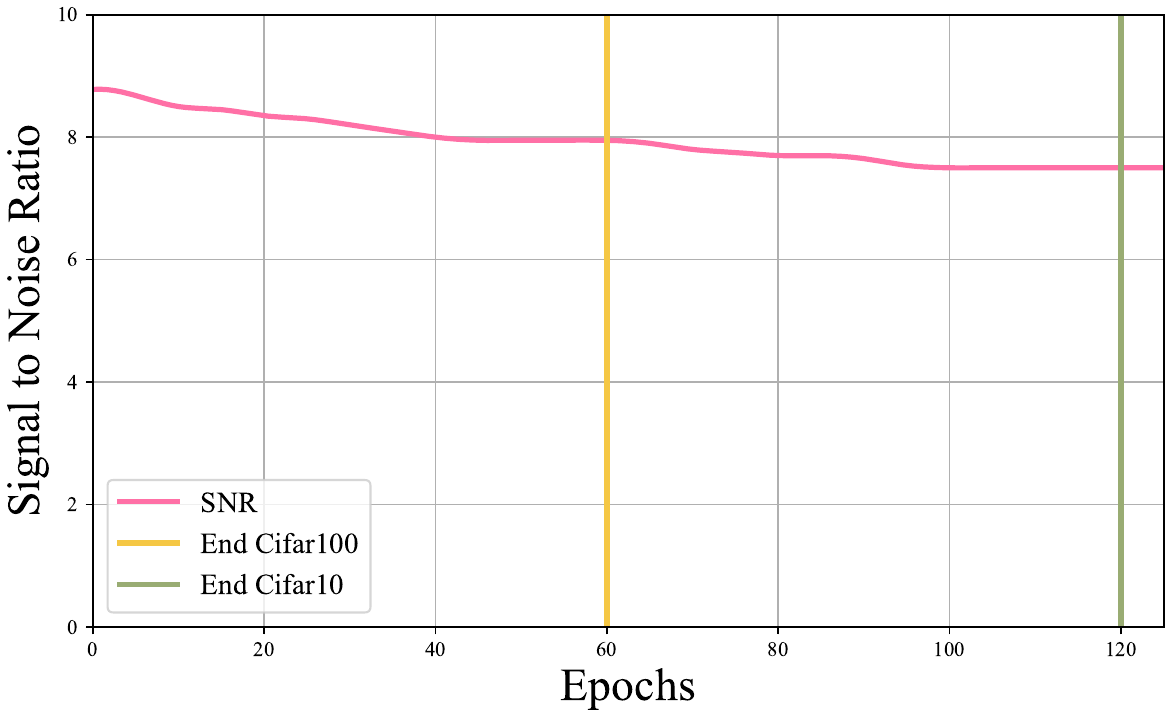}
             \caption{ResNet50 architecture trained on Imagenet dataset~\cite{imagenet_cvpr09} and first repurposed on CIFAR100 dataset and secondly repurposed on CIFAR10 dataset~\cite{krizhevsky2009learning}.}
             \label{fig:snr_resnet50_imagenet_cifar100_cifar10}
        \end{subfigure}
        \hfill
	    \begin{subfigure}{.32\textwidth}
            \centering
            \includegraphics[width=\columnwidth]{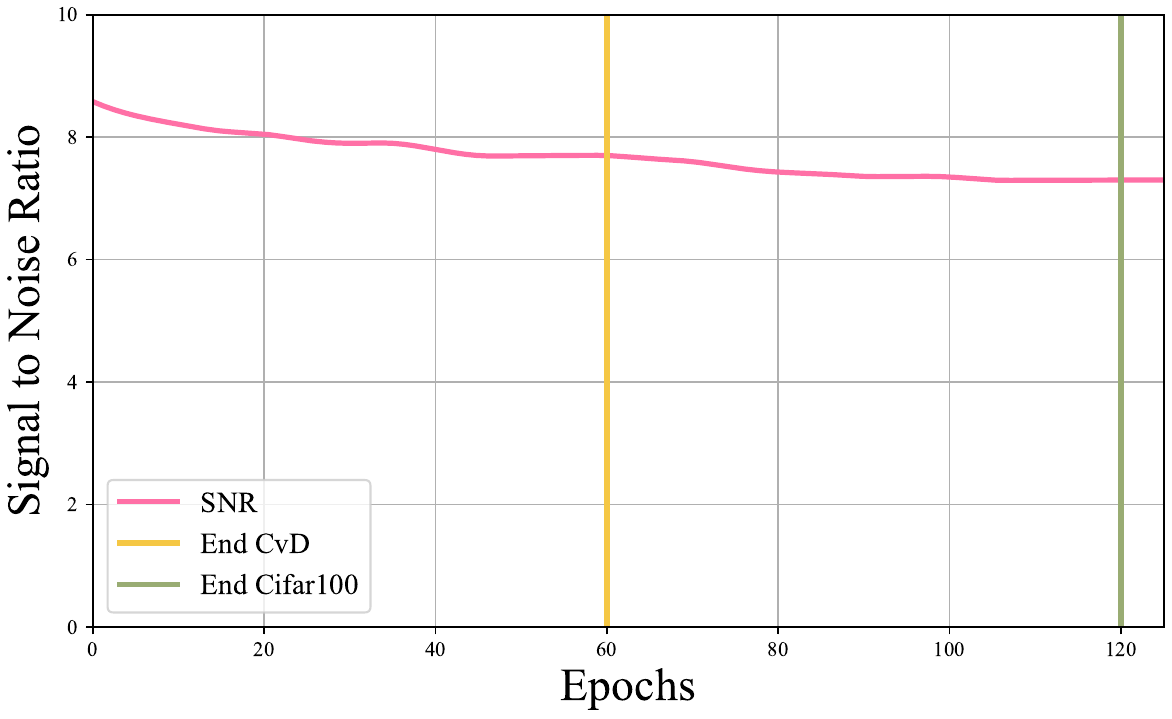}
             \caption{VGG11 architecture trained on Imagenet dataset, first repurposed on Cats vs. Dogs dataset, and secondly repurposed on CIFAR100 dataset~\cite{krizhevsky2009learning}.}
             \label{fig:snr_vgg11_imagenet_catsvdogs_cifar100}
        \end{subfigure}
            \hfill
	    \begin{subfigure}{.32\textwidth}
            \centering
            \includegraphics[width=\columnwidth]{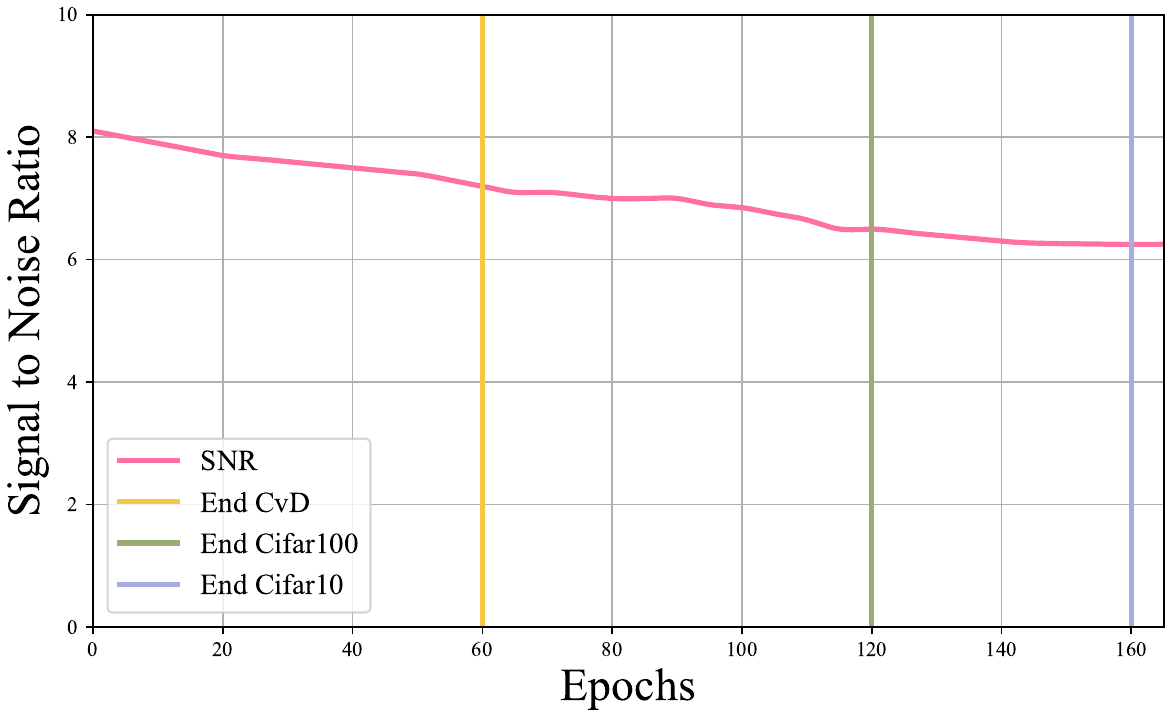}
             \caption{ResNet101 architecture trained on Imagenet dataset~\cite{imagenet_cvpr09} first repurposed on Cats vs. Dogs dataset, secondly on CIFAR100 and thirdly CIFAR10.}
             \label{fig:snr_resnet101_imagenet_catsvdogs_cifar100_cifar10}
        \end{subfigure}
        \bigskip 
        \caption{The effect of multiple re-purposing steps to the SNR of the malware payloads embedded in different DNN architectures via \name. The malware payload injected in all cases was Stuxnet.} 
    \label{fig:snr_finetuning_multistep}
\end{figure*}

The experimental evaluation in the previous sections focused on embedding malware payloads in standard, 32-bit floating point (FP32) network parameters. However, recent results show that half-precision (16-bit) floating point formats provide similar levels of performance as FP32 on the same architecture, while requiring much less memory and decreasing computation time~\cite{kalamkar2019study}. This section evaluates the performance of \name embedding technique in such half-precision architectures, which we refer to as \emph{reduced parameter width setting}. The reduction in the parameter bit-width is significant for payload embedding. Indeed, for each block of the payload, each CDMA-encoded bit is added to the same set of parameters of the model (see Section~\ref{app:implementation}). Therefore, successive approximations caused by the reduced precision can accumulate and affect the model parameters in unpredictable ways, leading to degradation in performance or even failure to embed the payload.

We test the effectiveness of \name in reduced parameter width settings by studying the embedding performance in the Llama2 model~\cite{touvron2023llama}, which uses the BFloat16 half-precision format to represent the model parameters. Furthermore, evaluating \name on the Llama2 model allows us to test the effects of our embedding technique on a considerably more complex architecture compared to previous experiments. Table~\ref{table:results_malware_injection_llama} shows the results of our analysis. We evaluate \name's performance penalty over multiple malware payloads using the MMLU dataset --- a standard dataset used in NLP to evaluate the performance of language models across a wide range of tasks. As highlighted in the table, the performance of \name models remains largely similar to the clean model's counterpart, with variations in the order of $\sim1$ percentage point --- well within the margin of error. Overall, these results demonstrate that \name's embedding technique is successful even when models employ reduced bit-width parameters. Moreover, the performance penalty induced by \name in the base model is minimal, further proving that assessing whether a model has been tampered with is challenging.

\begin{tcolorbox}\noindent
        \textbf{Remark:} Regarding the performance evaluation, we note that the \name models exhibit similar performance to their baseline counterparts in different scenarios such as standard, federated learning and even reduced parameter-width setting.
\end{tcolorbox}

\subsection{Robustness}\label{sec:robustness}

Pre-trained models obtained from online repositories are rarely used as-is: typically they need to be adapted for a specific domain or task before being employed. This section evaluates the robustness of \name against post-training processes that alter the parameters of the trained model. In particular, we analyze how fine-tuning the \name model affects the embedded payload. Fine-tuning is the process where part of the model parameters are updated in order to either improve the model on the current task, or re-purpose the model for a different (but similar) task. 
To measure the disruption that fine-tuning causes to the \name payload signal, we use the signal-to-noise ratio (SNR) metric, which indicates how strong the signal of our embedded payload is compared to noise, which in our case are the clean model parameters. Figure~\ref{fig:snr_finetuning} reports the SNR on different DNN architectures that are initially trained on one task, have had the malware payload embedded in them via \name, and then repurposed in a similar task as it would commonly happen in reality. The reported SNR is the ratio of signal power to noise power and is expressed in decibels (dB). Torrieri~\cite{spread_spectrum_principles} proves the threshold SNR values required for CDMA extraction to successfully separate (and thus decode) the wanted signal from the unwanted noise. When SNR is higher than 1:1 (greater than 0dB), there is more signal than noise, meaning that the payload can be extracted correctly. 

Figure~\ref{fig:snr_mnist_fashionmnist} shows the change in SNR for a ResNet50 model pre-trained on FashionMNIST~\cite{fashion_mnist} and fine-tuned on MNIST. The Stuxnet malware is injected into the pre-trained model via \name and, afterward, the model is fine-tuned to solve the MNIST digit recognition task. The figure plots the change in SNR of the payload with respect to the number of fine-tuning epochs. The fine-tuning is performed for the same number of epochs that were originally used to train on FashionMNIST. As we can see from the plots, the SNR of the \name payload initially decreases, but as the strength of the gradient updates decreases and the weights converge, the SNR plateaus.

Figure~\ref{fig:snr_cifar100_cifar100} plots the change in SNR for a VGG11 model pre-trained on CIFAR10 and fine-tuned on CIFAR100. The Stuxnet malware is injected into the pre-trained model via \name and the model is then fine-tuned to solve the CIFAR100 task. Similar to ResNet50, the SNR of the \name payload initially slightly decreases, but then stabilizes around $\sim7.5$dB. Such an SNR indicates that the payload signal in the model weights is still significantly strong, which allows for correct extraction.

Figure~\ref{fig:snr_imagenet_catsdogs} displays the change in SNR for a ResNet50 model pre-trained on Imagenet and fine-tuned on the Cats vs. Dogs task. The Stuxnet malware is injected into the pre-trained model via \name, and the model is then re-purposed to solve the Cats vs. Dogs task. As in the previous experiments, the SNR of the \name payload plateaus after an initial decrease. However, the payload signal remains significantly strong throughout the whole fine-tuning process, allowing for its correct extraction.

Figure~\ref{fig:snr_finetuning_multistep} illustrates the SNR changes in models undergoing multiple repurposing stages, simulating real-world fine-tuning scenarios. Specifically, Figure~\ref{fig:snr_resnet50_imagenet_cifar100_cifar10} shows a ResNet50 model with Stuxnet malware injected via \name, pre-trained on ImageNet, and then fine-tuned on CIFAR100 followed by CIFAR10. While the SNR initially drops due to strong gradient updates, it stabilizes as the model converges. This pattern holds across both fine-tuning phases, with the \name payload remaining intact and fully recoverable. An initial takeaway is that SNR degradation is less pronounced when fine-tuning on simpler tasks, as observed in the transition from CIFAR100 (a more complex task) to CIFAR10 (a simpler one)

Figure~\ref{fig:snr_vgg11_imagenet_catsvdogs_cifar100} illustrates the evolution of the SNR for a VGG11 model pre-trained on the ImageNet dataset and subsequently embedded with the Stuxnet malware using \name. The model was fine-tuned in two successive stages: first on the Cats vs. Dogs dataset, and then on CIFAR100. The plot shows that the SNR of the embedded \name payload initially declines during fine-tuning, reflecting the effect of strong early gradient updates. However, as training progresses and the model weights converge, the rate of SNR loss slows and eventually plateaus. This pattern remains consistent even after the second repurposing—from Cats vs. Dogs to CIFAR100—where a minor drop in SNR is again followed by rapid stabilization. Importantly, despite both repurposing phases, the \name payload remains fully embedded within the model and can be reliably retrieved.

Figure~\ref{fig:snr_resnet101_imagenet_catsvdogs_cifar100_cifar10} depicts the evolution of the SNR in a ResNet101 model pre-trained on the ImageNet dataset, into which the Stuxnet malware was embedded using \name. The model underwent a prolonged fine-tuning pipeline across three stages: first on the Cats vs. Dogs dataset, then on CIFAR100, and finally on CIFAR10. As observed in previous experiments, the SNR of the embedded \name payload declines initially due to substantial gradient updates but gradually stabilizes as the model converges. This plateauing trend persists across all three repurposing stages, including the final adaptation to CIFAR10. Notably, despite the extended sequence of fine-tuning tasks, the \name payload remains fully embedded in the model and can be reliably recovered in its entirety.

These findings reinforce that fine-tuning—even across multiple stages—does not significantly deteriorate the \name payload signal. In each repurposing step, a moderate drop in SNR is observed, typically more noticeable when fine-tuning on complex tasks, which induce larger changes in model parameters to achieve convergence. Despite this, the overall reduction remains limited after each stage. In the analyzed scenario, where the Stuxnet payload was embedded via \name with an initial SNR slightly above 8:1, the signal remains robust and fully recoverable. Importantly, reducing the SNR to levels where payload bits become unreliable would require a large number of diverse, successive repurposings.

\begin{tcolorbox}\noindent
        \textbf{Remark:} Regarding the robustness evaluation, we note that the malware payload embedded via \name can not be hampered even when the \name models undergo different fine-tuning and repurposing stages, always exhibiting an SNR ratio above 1:1—ensuring that the embedded signal can still be distinguished from background noise and accurately decoded.
\end{tcolorbox}
\section{Discussion and Countermeasures}\label{sec:defenses}

This section discusses hyperparameter trade-offs and explores potential mitigation strategies against payload embedding in DNNs. We categorize potential defenses based on their applicability to traditional versus collaborative learning settings, noting that \name is the only known attack currently feasible in the collaborative setting, as prior methods are not applicable in that context. Furthermore, since the mitigation strategies evaluated in section~\ref{sec:evaluation} successfully neutralized threate from prior attacks such as~\cite{stegonet,evilmodel_1,evilmodel_2}, our focus here is on strategies that could potentially mitigate \name as well as earlier techniques.

\subsection{Hyperparamenters Tradeoffs}
There are three primary hyperparameters that influence \name's effectiveness, stealthiness, and robustness: (1) the $\gamma$ factor; (2) the spreading code size; and (3) the number of LDPC redundancy bits. The $\gamma$ factor controls the strength of the malware payload signal that is injected in the network weights, directly impacting both the effectiveness of \name's encoding and its stealthiness. A higher $\gamma$ improves the signal-to-noise ratio (SNR), making the encoded payload more distinguishable from the noise (i.e., the clean model's weights) and easier to correctly extract later. However, increasing $\gamma$ also reduces the stealthiness of the embedding, as it amplifies the magnitude of weight changes introduced during injection. By making certain assumptions about the distribution of the noise (e.g., Gaussian-distributed), one can mathematically derive a $\gamma$ factor that guarantees sufficient SNR for reliable payload extraction while maintaining undetectability due to the inherent noise of the channel~\cite{Verdu98,spread_spectrum_principles}. In practice, accurately modeling the noise distribution is challenging and an empirical approach is required to identify an appropriate $\gamma$ to ensure reliable extraction. The spreading code size used during the CDMA encoding process directly impacts the robustness of the payload to model modifications and the stealthiness of the injection with respect to final model performance. Larger spreading codes introduce greater encoding redundancy, enhancing resilience to subsequent changes in the model's weights. However, they also require altering a larger number of model parameters, which may adversely affect the model's overall performance. Finally, the number of LDPC redundancy bits influences both the robustness of the payload to weight changes and the stealthiness of the injection. LDPC enables correction of extraction errors introduced by additional noise in the model's weights. However, the LDPC auxiliary information must also be CDMA-encoded and injected into the model. Consequently, using excessively large LDPC redundancy can degrade the model's performance and increase the risk of detection.

\setlength{\tabcolsep}{4.5pt} 
\begin{table}[t]
\centering
\caption{Comparison of the effects of different levels of model parameter pruning on Stuxnet malware payload injected into ResNet50 using Liu et al.~\cite{stegonet}, Wang et al.~\cite{evilmodel_1}, Wang et al.~\cite{evilmodel_2} and \name malware embedding techniques.}
\begin{tabular}{c c c c c}
\toprule
& \multicolumn{4}{c}{\bfseries Does the payload survive?} \\
\cmidrule(l){2-5}
 P. Ratio &Liu et al.~\cite{stegonet} &Wang et al.~\cite{evilmodel_1} &Wang et al.~\cite{evilmodel_2} &Ours \\
\midrule
$25\%$ & \xmark & \xmark & \xmark & \cmark  \\
$50\%$ & \xmark & \xmark & \xmark & \cmark  \\
$75\%$ & \xmark & \xmark & \xmark & \cmark  \\
$90\%$ & \xmark & \xmark & \xmark & \xmark   \\
$99\%$ & \xmark & \xmark & \xmark & \xmark   \\
\bottomrule
\end{tabular}\label{table:pruning_outcome}
\end{table}

\subsection{Traditional setting countermeasures}
\subsubsection{Pruning}
Over the years, several approaches have been proposed to mitigate DNN attacks based on weight manipulation such as backdooring~\cite{li2021invisible}. One of the best-known mitigation techniques proposed is parameter pruning~\cite{badnets,removebackdoors}. The idea behind parameter pruning is that, since backdoors maliciously alter some parameters of the network such that an adversary can obtain a predefined output given an input with a specific trigger, it is possible to neutralize the attack if we can remove (prune) the malicious weights from the model. This is typically achieved by heavily altering or zeroing a subset of the model's parameters. A parallel can be drawn between backdooring attacks and malware injection in a DNN, since both approaches typically require the maliciously altered weights to remain unaltered for the attack to be successful. Consequently, parameter pruning is an interesting candidate for a potential defense against \name.
In prior payload embedding techniques, such as Liu et al.~\cite{stegonet}, each bit of the payload is always embedded in a single parameter of the model. Therefore, zeroing even one of the parameters of the network where one of the malicious payload bits are mapped results in a corrupted payload, defeating the attack. Typically, parameter pruning defenses zero several model parameters, thus making it highly likely that at least a subset of the maliciously modified parameters is pruned, and the payload corrupted.
Similar limitations affect~\cite{evilmodel_1,evilmodel_2}, where zeroing a single model parameter where part of the payload is mapped results in a corrupted payload.
Contrary to previous work, \name is highly resilient against parameter pruning defenses. The use of CDMA to spread each bit of the payload across a large number of parameters, coupled with the use of error-correcting LDPC codes makes \name robust against a large number of modifications to the parameters, including zeroing. Indeed, using \name it is sufficient that only a small subset of the maliciously altered parameters survives pruning for the payload signal to survive and allow successful CDMA decoding. Table~\ref{table:pruning_outcome} compares the effect of parameter pruning on \name against prior art. We train a ResNet50 model on the Imagenet dataset and embed the Stuxnet malware in the trained model using~\cite{stegonet,evilmodel_1,evilmodel_2}, and \name. 

Another form of model parameter pruning involves removing entire neurons, effectively altering the network architecture. These architectural pruning techniques are effective in mitigating \name and prior work~\cite{stegonet,evilmodel_1,evilmodel_2}. This is due to the payload extraction procedure’s reliance on the specific ordering of model parameters. While such methods can be effective against this class of attacks, they require significant machine learning expertise and access to a dataset that closely matches the distribution of the original training data. As a result, only well-resourced entities with the necessary technical knowledge and data access can reliably apply these defenses without compromising model functionality. 
Furthermore, an adaptive adversary could preemptively apply pruning during model development before embedding the malicious payload using \name. Since further pruning of an already pruned model is likely to severely degrade performance, this strategy would limit the defender's ability to leverage pruning as a mitigation strategy.

\subsubsection{Fine-tuning}
Another potential mitigation against payload embedding is fine-tuning--- a common approach to repurpose pretrained models to new data distributions and tasks. Fine-tuning alters the model parameters, even extensively, and could in principle degrade or remove embedded payloads. In Table~\ref{table:regular_finetuning_outcome}, we report and compare the resilience between \name and prior work under fine-tuning. As illustrated in the table, prior work~\cite{stegonet,evilmodel_1,evilmodel_2} is easily disrupted even by limited fine-tuning, whereas \name payloads remain intact. This robustness is due to \name's use of CDMA to spread each bit of the payload across a large number of network parameters, combined with LDPC error-correcting codes, which ensure resilience against substantial modifications to the parameters, including those introduced by fine-tuning.

\begingroup
\setlength{\tabcolsep}{4.5pt} 
\begin{table}[t]
\centering
\caption{Comparison of the effects of different fine-tuning epochs on Stuxnet malware payload injected into ResNet50 using Liu et al.~\cite{stegonet}, Wang et al.~\cite{evilmodel_1}, Wang et al.~\cite{evilmodel_2} and \name malware embedding techniques.}
\begin{tabular}{c c c c c}
\toprule
& \multicolumn{4}{c}{\bfseries Does the payload survive?} \\
\cmidrule(l){2-5}
 Epochs &Liu et al.~\cite{stegonet} &Wang et al.~\cite{evilmodel_1} &Wang et al.~\cite{evilmodel_2} &Ours \\
\midrule
10 & \xmark & \xmark & \xmark & \cmark  \\
50 & \xmark & \xmark & \xmark & \cmark  \\
100 & \xmark & \xmark & \xmark & \cmark   \\
150 & \xmark & \xmark & \xmark & \cmark   \\
200 & \xmark & \xmark & \xmark & \cmark   \\
\bottomrule
\end{tabular}\label{table:regular_finetuning_outcome}
\end{table}
\endgroup

\subsubsection{Compression}
A more sophisticated family of techniques that alter the weights of the network itself is model compression. As the name implies, the goal of model compression techniques is to reduce the computational and memory footprint of modern DNNs, allowing their execution on commodity devices. The most well-know and widespread model compression technique is model quantization~\cite{xiao2023smoothquant}, which reduces the bit-width of the parameters used in the DNN. While model quantization can successfully mitigate prior embedding approaches and \name if it is applied after the payload injection, Section~\ref{sec:llama_eval} shows that \name is effective when bit-width reduction is applied before the payload is embedded. Given the complexity that quantizing a DNN entails, it is unlikely that DNN consumers would possess the technical skills and know-how required for the quantization. It is far more likely that DNN consumers would look for an already quantized model to download and utilize~\cite{llama_quantized}. 

\subsubsection{Safe Serialization}
Another potential avenue for mitigating payload injection attacks is to target the extraction and self-execution stages. As discussed in Section~\ref{sec:trigger_mechanisms}, one method of rendering \name self-executable involves exploiting unsafe deserialization. Specifically, adversaries can modify a model's saved file to inject additional extraction code, such that invoking deserialization functions triggers its execution, assembling and executing the malicious payload embedded in the parameters. Self-extraction and execution can be mitigated by adopting safe deserialization functions. One such recently proposed approach is SafeTensors\footnote{\url{https://huggingface.co/docs/safetensors/}}. SafeTensors is a binary tensor storage format specifically designed to provide safe, efficient, and fast storage and loading of tensor data. Unlike traditional pickle-based formats (e.g., PyTorch \textit{.pt} files), SafeTensors prevents arbitrary code execution when loading data, reducing the risk of malicious payload injection. As a result, many deep learning frameworks--- including Hugging Face Transformers---have adopted SafeTensors to securely distribute models and safeguard production environments from deserialization-based attacks. However, adversaries can still leverage payload embedding techniques to covertly bypass detection engines, and subsequently rely on separate executables or vulnerabilities to extract and execute the hidden payload.

\subsection{Collaborative Learning setting countermeasures}

\subsubsection{Federated Learning}
Over the years, various distributed machine learning techniques have been introduced to enhance privacy and efficiency by decentralizing model training, with federated learning emerging as the predominant technique due to its efficiency and ease of integration. Section~\ref{sec:evaluation_fl} demonstrated that an adversary operating within the federated learning paradigm can successfully embed arbitrary payloads into the weights of the global model, which is then shared among all participants. While this behavior significantly amplifies the threat posed by \name in federated settings, a key limitation of the attack in this scenario is that the adversary must first gain a foothold in the target devices to trigger the malicious payload. In contrast, in the traditional setting the adversary has more opportunities to exploit library vulnerabilities enabling the creation of fully self-executing stegomalware, where the payload is extracted and executed as soon as the model is loaded. Nevertheless, the ability to embed arbitrary payloads enable adversaries to pursue a range of threat scenarios. Adversaries can leverage seemingly benign applications capable of extracting and executing the malware without containing overtly malicious code or downloaders, instead performing innocuous-looking mathematical operations to recover the payload from the DNN.

In response to such threats, various defense mechanisms can be implemented by one or more entities involved in the collaborative learning scheme. One such approach is architectural pruning. As previously discussed, architectural pruning enables the removal of redundant neurons from the trained DNN without significantly compromising its performance. In this context, the global parameter server can proactively apply pruning after the final training iteration before distributing the model to participants. Since \name's injection and extraction procedures rely on specific weight ordering, altering the model structure effectively prevents successful extraction. Similarly, model compression through quantization can also be employed to eliminate the malicious payload from the final trained model. However, this defense does not fully eliminate the risk, as the adversary may extract the payload at any point before the final model is pruned and distributed to users. 

\subsubsection{Split Learning}
Another approach that has emerged in recent years aiming to address some of the privacy concerns in federated learning is split learning (SL)~\cite{splitlearning}. SL partitions a deep learning model into multiple segments across different participants. Unlike traditional federated learning where each participant trains a complete local model, SL requires participants to trains only a portion of the model. Inference and backpropagation are performed by exchanging intermediate activations and partial gradient information between the participants and the central server. This approach, on top of the benefits offered by federated learning, also reduces computational burden on client devices, as part of the model layers are offloaded to the central server. However SL presents certain drawbacks. The frequent exchange of activations and gradients introduces communication overhead and necessitates synchronization between participants. This challenges are less pronounced in federated learning, which typically requires fewer communication rounds and looser coordination. Consequently, SL has been primarily adopted in domains with stringent privacy concerns---such as medical data---while seeing more limited adoption in less privacy-sensitive applications. 

Concerning payload injection, SL is inherently robust against all techniques that rely on direct weight or gradient manipulation, including \name. This robustness stems from the fact that participants never share model weights or gradient updates, but just activation outputs. In the FL setting, \name embeds the payload directly into the gradient updates that are sent to the global parameter server for aggregation, allowing it to survive the averaging process. In SL, since that no gradients (weights) are exchanged among participants, such approach is unfeasible in its current state. Despite SL’s robustness to this class of attacks, federated learning remains the more widely adopted paradigm due to its simpler coordination, faster convergence, and reduced complexity in collaboratively training models with privacy considerations. 

\subsubsection*{Takeaway}
The evolving threat landscape for \name and payload embedding remains a significant concern, highlighting the need for continued research and proactive security measures. Consequently, future work should be dedicated to developing robust defense mechanisms to mitigate attacks such as \name, ensuring resilience even within the federated learning paradigm. 
\section{Related Work}\label{sec:related_work}

Liu et al.~\cite{stegonet}, to the best of our knowledge, are the first to create a new breed of stegomalware through embedding malware into a deep neural network model. They proposed four methods to embed a malware binary into the model and designed and evaluated triggering mechanisms based on the logits of the model. The first malware embedding method introduced by Liu et al.~\cite{stegonet} is LSB substitution, where the malware bits are embedded into the model by replacing the least significant bits of the models' parameters. 
The second method consists of a more complex version of the LSB substitution. The idea behind it resides in substituting the bytes of a set of models' weight parameters with the bytes of the malware payload. After that, they perform model retraining by freezing the modified weight parameters (where the malware is placed) to restore model performance using only the remainder of the weight parameters.
Alongside those two methods, Liu et al.~\cite{stegonet} proposed mapping-based techniques to map (and sometimes substitute) the values or the sign of the network's weights to the bit of the malware. They call these methods value-mapping and sign-mapping, respectively.

Liu et al.~\cite{stegonet} demonstrated that, in the case of LSB substitution, resilience training, and value-mapping, even a single flip in one bit would corrupt the malware, thus rendering these payload embedding techniques unreliable and unusable in practice since even a simple fine-tuning could disrupt the malware extraction.
Sign-mapping is the most robust of the four payload embedding techniques proposed in~\cite{stegonet}, but it suffers from several limitations. Sign-mapping maps the bits of the malware payload to the sign of the model's weights. This means that the number of bits it can map is more limited than other methods. Based on data reported by Liu et al.~\cite{stegonet}, the amount of bits that the sign-mapping technique can embed is of the same order of magnitude as the number of bits \name can embed. 
Compared to \name, sign-mapping of Liu et al.~\cite{stegonet} requires significantly more information to perform the payload extraction, i.e., the permutation map of the weights. In contrast, \name extractor only needs to know the seed to generate CDMA spreading codes. 

Wang et al.~\cite{evilmodel_1} proposed fast-substitution as a way to embed malware into a deep neural network. Fast-substitution works by substituting the bits of a selected group of weights in a model with the ones from the malware. In case the performance of the resulted model is highly impaired, the authors restore the performance of the model similar to the resilience training method presented by Liu et al.~\cite{stegonet}, i.e., freezing the group of weights selected to embed the payload and retrain the model. Fast-substitution suffers the same drawbacks as LSB substitution, resilience training, and value-mapping from Liu et al.~\cite{stegonet}, making it unusable in the supply-chain attack scenario, where the models are usually fine-tuned. 
Wang et al. further extended their work~\cite{evilmodel_2} proposing two additional techniques: most significant byte (MSB) reservation and half-substitution. Both the methods rely on the fact that the model performance is better maintained when the first bytes of each model weight are preserved. Even if they can guarantee less performance degradation, those methods suffer from the same weaknesses as fast-substitution, making them unsuitable in cases where fine-tuning is used.
\section{Ethical discussion}\label{sec:ethical}
The ever-growing integration of machine learning solutions across multiple domains presents a broad landscape that adversaries can exploit for malicious activities. The intent of this work is not to provide malicious actors with innovative techniques for creating stegomalware and potentially cause harm.
Instead, the purpose here is to raise awareness regarding the existence and accompanying risks of such potential attack vector.
We encourage consumers of machine learning-based solutions to obtain services from reputable and trustworthy sources. Additionally, we aspire to motivate researchers and vendors to develop resilient solutions that preemptively address and mitigate these kind of threats.
\section{Conclusions}\label{sec:conclusions}
We proposed \name, a novel malware embedding technique for DNN based on CMDA and LDPC. Through extensive empirical analysis, we showed that \name models incur little to no performance penalty, that \name generalizes across a wide variety of architectures and datasets, and that state-of-the-art malware detection and statistical analysis techniques fail to detect the malware payload. We implemented a self-extracting and self-executing malware proof of concept, showing the practicality of the attack against a widely adopted DNN framework. We demonstrated that \name is effective even in distributed learning settings such as federated learning, and that it is not affected by reduced parameter bit-width architectures. Finally, we proved that \name is resilient to parameter manipulation techniques, as well as potential defenses such as parameter pruning. 

Our work highlights a novel, undetectable threat to the ML supply chain. We aim to increase awareness of these family of attacks among both the research community and industry, and to spur further research in mitigation techniques to address this threat. 

\section*{Acknowledgements} This work was partially supported by project SERICS (PE00000014) under the NRRP MUR program funded by the EU-NextGenerationEU and by Gen4olive, a project that has received funding from the European Union’s Horizon 2020 research and innovation programme under grant agreement No. 101000427.

\bibliographystyle{IEEEtranS}
\bibliography{bibliography}
\vspace{-1em}
\begin{IEEEbiography}[{\includegraphics[width=1in,height=1.25in,clip,keepaspectratio]{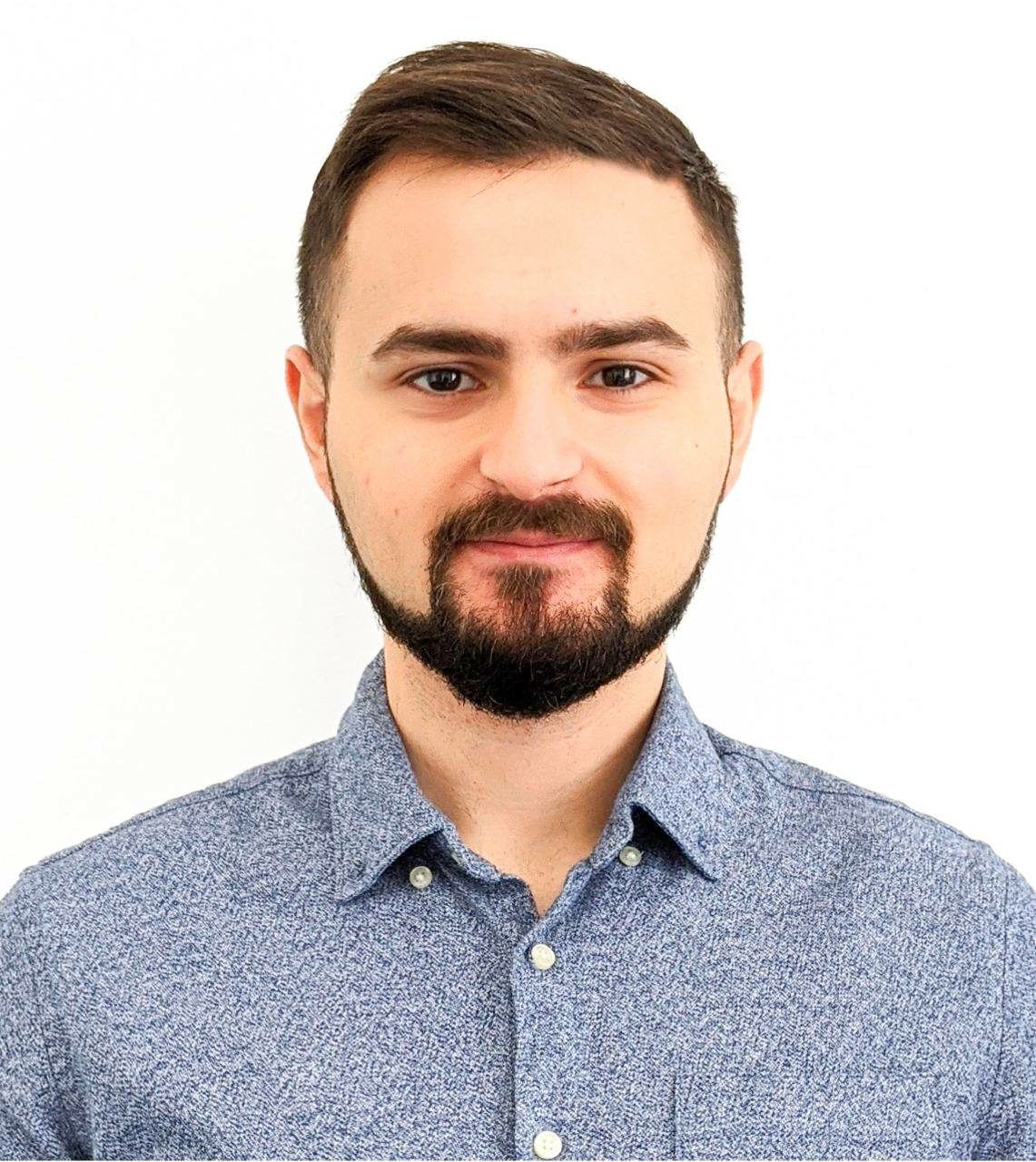}}]
{Dorjan Hitaj} is an Assistant Professor with the Computer Science Department of Sapienza University of Rome. His research focus lies in the intersection between machine learning and security. He obtained his Ph.D. in Computer Science with high honors from Sapienza University of Rome in 2022.
\end{IEEEbiography}

\begin{IEEEbiography}[{\includegraphics[width=1in,height=1.25in,clip,keepaspectratio]{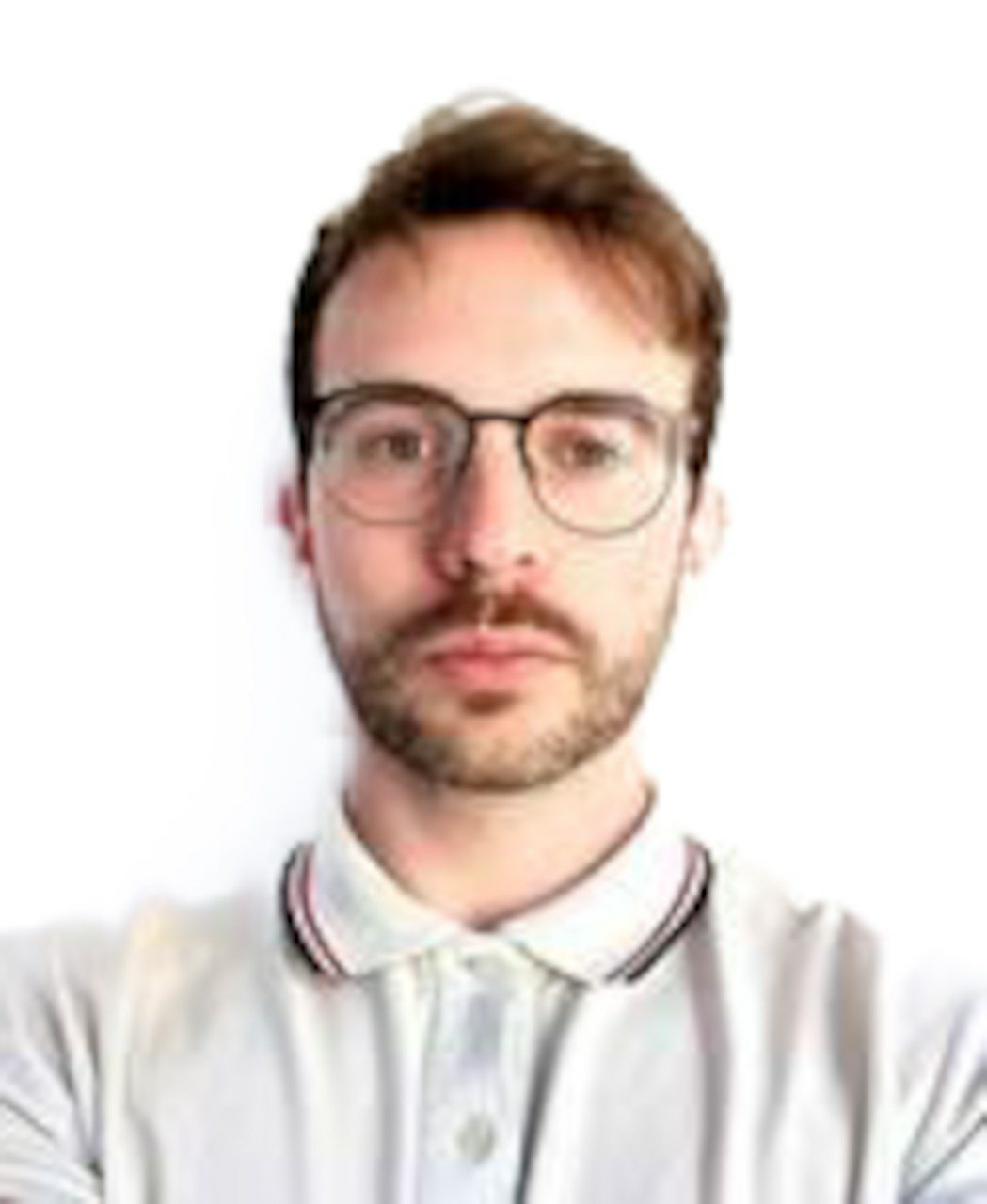}}]
{Giulio Pagnotta} is a Postdoctoral researcher in Computer Science at Sapienza University of Rome. His research focus is on machine learning and security. He obtained his Ph.D. in Computer Science with high honors from Sapienza University of Rome in 2023.
\end{IEEEbiography}

\begin{IEEEbiography}[{\includegraphics[width=1in,height=1.25in,clip,keepaspectratio]{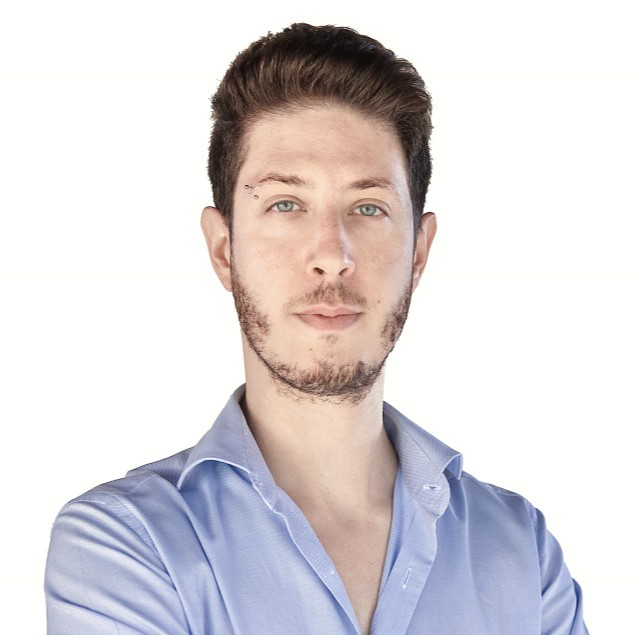}}]
{Fabio De Gaspari} is an Assistant Professor with the Computer Science Department of Sapienza University of Rome. His research focus is on machine learning, cybersecurity and computer networks. He obtained his Ph.D. in Computer Science from Sapienza University of Rome in 2019.
\end{IEEEbiography}

\vspace{-1em}
\begin{IEEEbiography}[{\includegraphics[width=1in,height=1.25in,clip,keepaspectratio]{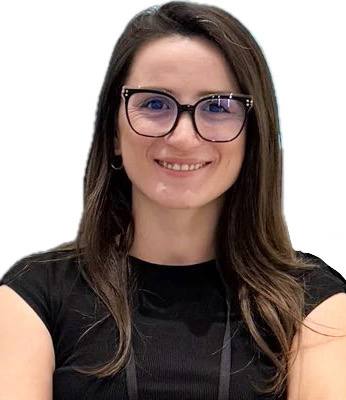}}]
{Sediola Ruko} is a PhD student with the DEIM department of Tuscia University. Her research focuses on applied machine learning and digital mapping technologies, with particular emphasis on their integration to address real-world challenges in urban navigation, accessibility, and personalized route planning.
\end{IEEEbiography}

\vspace{-1em}
\begin{IEEEbiography}[{\includegraphics[width=1in,height=1.25in,clip,keepaspectratio]{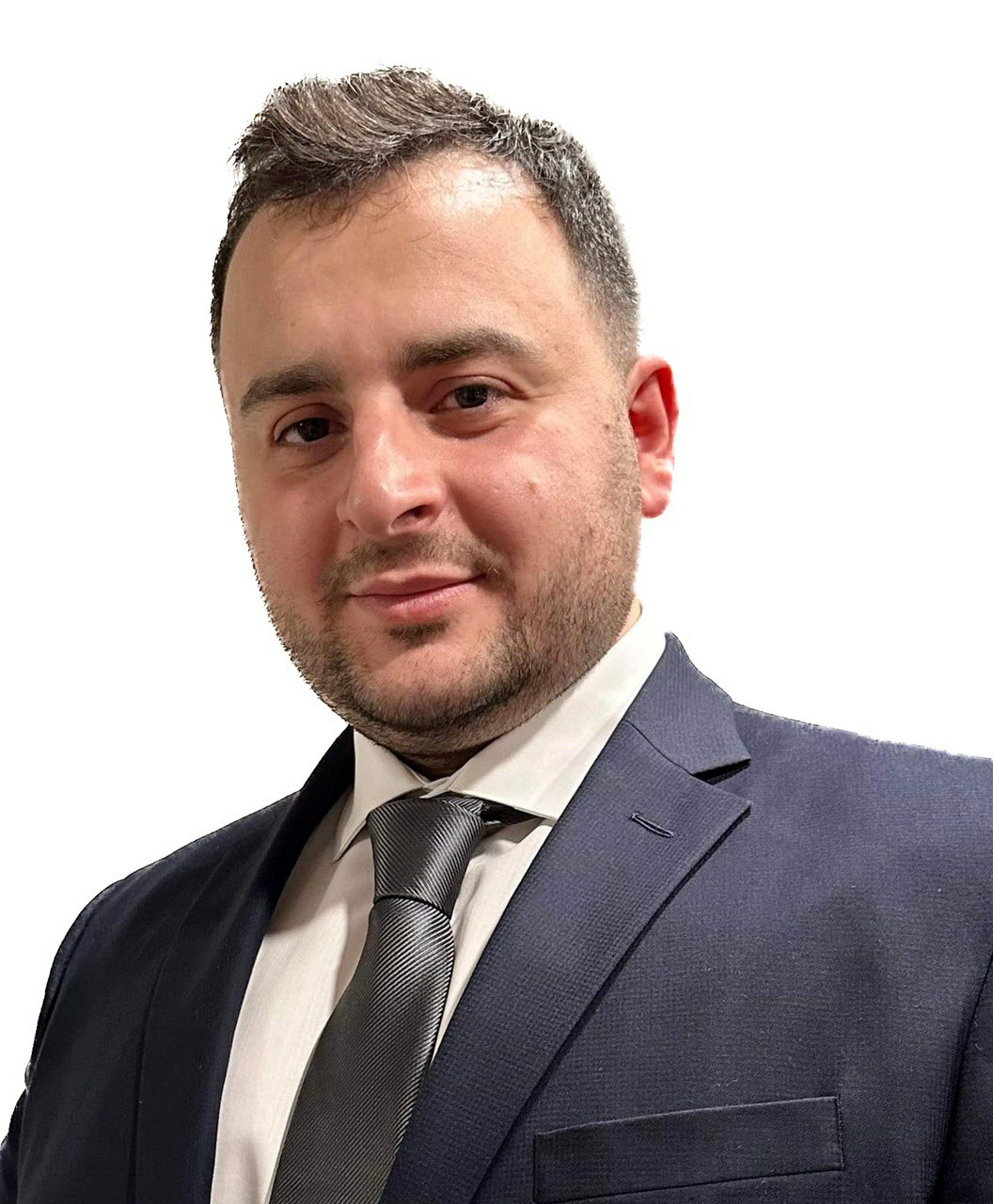}}]
{Briland Hitaj} is an Advanced Computer Scientist at SRI International. His expertise is in security and privacy in deep learning systems. Prior to joining SRI International, he was a Visiting Research Scholar at Stevens Institute of Technology. Dr. Hitaj obtained his Ph.D. in Computer Science from Sapienza University of Rome in 2018. 
\end{IEEEbiography}

\vspace{-1em}
\begin{IEEEbiography}[{\includegraphics[width=1in,height=1.25in,clip,keepaspectratio]{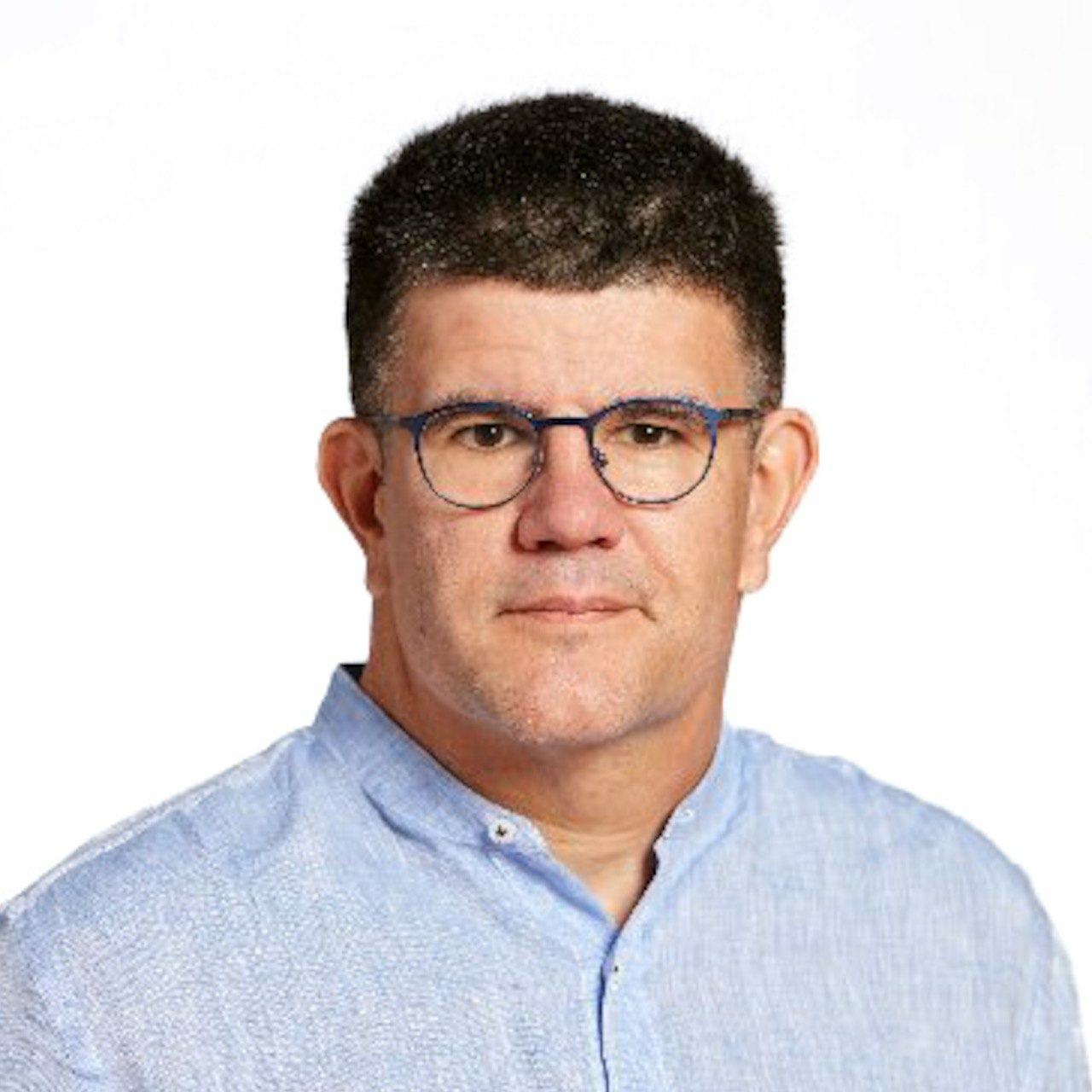}}]
{Fernando Perez-Cruz}
received a Ph.D. in Electrical Engineering from the Technical University of Madrid and has held roles at Bell Labs, Amazon, and the Swiss Data Science Center. Fernando was a visiting professor at Princeton, an associate professor at University Carlos III, and worked at the Gatsby Unit, Max Planck Institute, and BioWulf Technologies. Since 2024, he is a senior adviser at the Bank for International Settlements.
\end{IEEEbiography}

\vspace{-1em}
\begin{IEEEbiography}[{\includegraphics[width=1in,height=1.25in,clip,keepaspectratio]{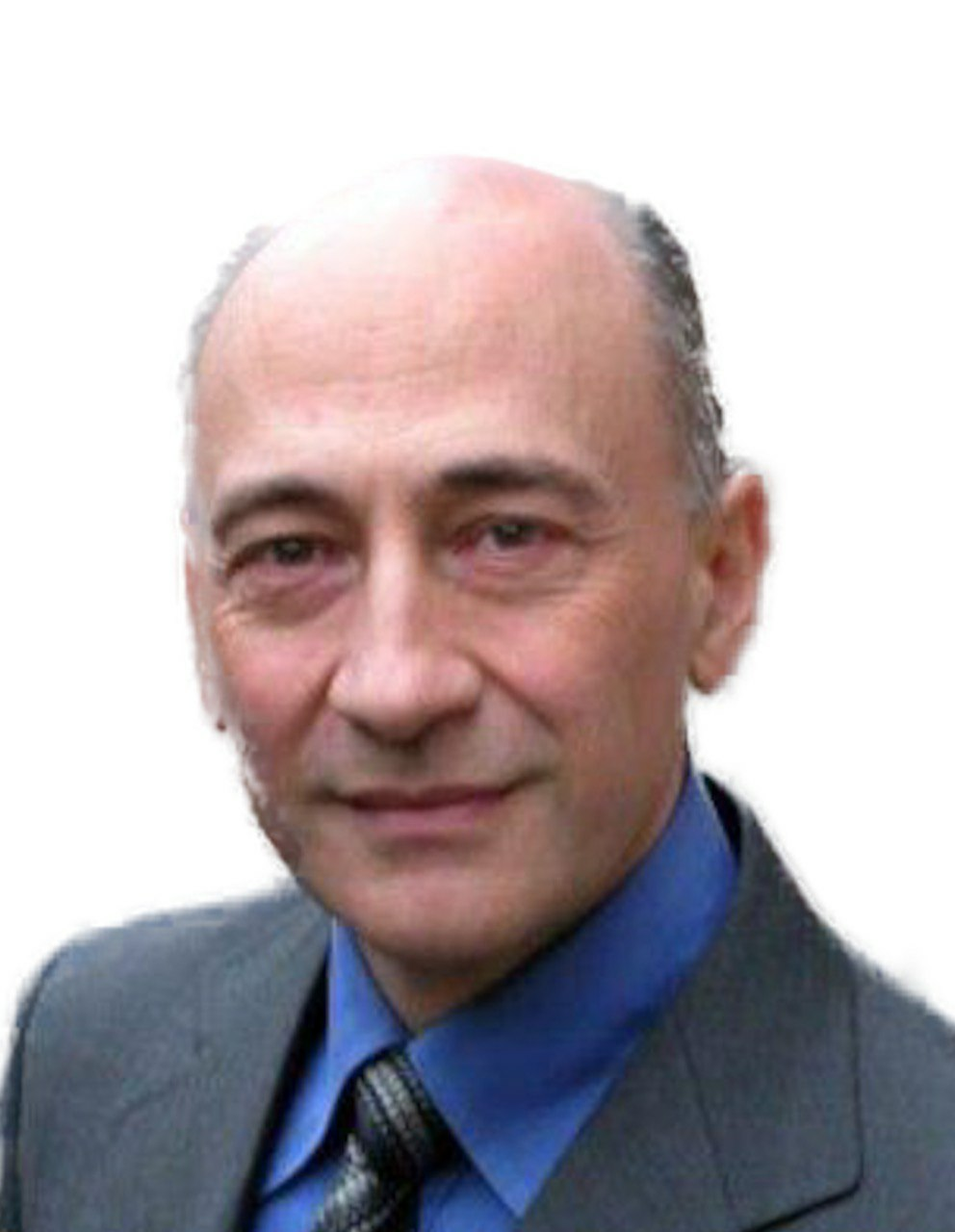}}]
{Luigi V. Mancini}
 is a Full Professor with the Computer Science Department of Sapienza University of Rome. He has authored over 130 scientific papers in international conferences and journals. He obtained his Ph.D. degree in Computer Science from the University of Newcastle, U.K., in 1989.
\end{IEEEbiography}
\end{document}